\documentclass[lettersize,journal]{IEEEtran}
\usepackage{amsmath,amsfonts}
\usepackage{algorithmic}
\usepackage{algorithm}
\usepackage{array}
\usepackage[caption=false,font=normalsize,labelfont=sf,textfont=sf]{subfig}
\usepackage{textcomp}
\usepackage{stfloats}
\usepackage{url}
\usepackage{verbatim}
\usepackage{graphicx}
\usepackage{cite}
\usepackage{amssymb}
\usepackage{epsfig}
\usepackage{caption}
\usepackage{textcomp}

\usepackage{pifont}

\usepackage{subfig} 
\usepackage{subfloat}

\usepackage{enumerate}
\usepackage{threeparttable}
\usepackage{booktabs} 
\usepackage{bm}

\usepackage{multirow} 
\usepackage{makecell}
\usepackage{setspace}
\newtheorem{assumption}{\textbf{Assumption}}
\newtheorem{definition}{\textbf{Definition}}
\newtheorem{lemma}{\textbf{Lemma}}
\newtheorem{Theorem}{\textbf{Theorem}}

\newtheorem{remark}{\textbf{Remark}}

\begin{document}
	
	\title{Communication-Efficient Distributed Learning via Sparse and Adaptive Stochastic Gradient}
	
	
	\author{Xiaoge Deng, Dongsheng Li, Tao Sun and Xicheng Lu
		\thanks{Xiaoge Deng, Dongsheng Li, Tao Sun and Xicheng Lu are with College of Computer, National University of Defense Technology, Changsha, Hunan, China.
			E-mail: dengxg@nudt.edu.cn, dsli@nudt.edu.cn, nudtsuntao@163.com}
		\thanks{Corresponding authors: Dongsheng Li and Tao Sun.}
		\thanks{Manuscript received April 19, 2021; revised August 16, 2021.}}
	
	\markboth{Journal of \LaTeX\ Class Files,~Vol.~14, No.~8, August~2021}%
	{Shell \MakeLowercase{\textit{et al.}}: A Sample Article Using IEEEtran.cls for IEEE Journals}
	
	\IEEEpubid{0000--0000/00\$00.00~\copyright~2021 IEEE}
	
	\maketitle
	
	\begin{abstract}
		Gradient-based optimization methods implemented on distributed computing architectures are increasingly used to tackle large-scale machine learning applications. A key bottleneck in such distributed systems is the high communication overhead for exchanging information, such as stochastic gradients, between workers. The inherent causes of this bottleneck are the frequent communication rounds and the full model gradient transmission in every round. In this study, we present SASG, a communication-efficient distributed algorithm that enjoys the advantages of sparse communication and adaptive aggregated stochastic gradients. By dynamically determining the workers who need to communicate through an adaptive aggregation rule and sparsifying the transmitted information, the SASG algorithm reduces both the overhead of communication rounds and the number of communication bits in the distributed system. For the theoretical analysis, we introduce an important auxiliary variable and define a new Lyapunov function to prove that the communication-efficient algorithm is convergent. The convergence result is identical to the sublinear rate of stochastic gradient descent, and our result also reveals that SASG scales well with the number of distributed workers. Finally, experiments on training deep neural networks demonstrate that the proposed algorithm can significantly reduce communication overhead compared to previous methods.
	\end{abstract}
	
	
	\section{Introduction}
	\label{sec:intro}
	\IEEEPARstart{O}{ver} the past few decades, the scale and complexity of machine learning (ML) models and datasets have significantly increased \cite{brown2020language,deng2009imagenet}, leading to greater computational demands and spurring the development of distributed training \cite{dean2012large,li2014scaling,xing2015petuum}. A large number of distributed machine learning tasks can be described as
	\begin{equation}
		\label{eqn_1}
		\min_{\omega \in \mathbb{R}^{d}} ~ F(\omega):=\frac{1}{M}\sum_{m \in \mathcal{M}} \mathbb{E}_{\xi_{m} \sim \mathcal{D}_{m}}\left[f_{m}(\omega; \xi_{m})\right],
	\end{equation}
	where $\omega$ represents the parameter vector to be learned, $d$ is the dimension of parameters, and $\mathcal{M}:=\{1, \dots, M\}$ denotes the set of distributed workers. $\{f_{m}\}_{m=1}^{M}$ are smooth (not necessarily convex) loss functions kept at worker $m$, and $\{\xi_{m}\}_{m=1}^{M}$ are independent random data samples associated with probability distribution $\{\mathcal{D}_{m}\}_{m=1}^{M}$. For simplicity, we define $F_{m}(\omega):=\mathbb{E}_{\xi_{m} \sim \mathcal{D}_{m}}\left[f_{m}(\omega; \xi_{m})\right]$. Let $\omega^{*}$ be the optimal solution and $F^{*}:=F(\omega^{*})$. Problem \eqref{eqn_1} is encountered in a broad range of distributed machine learning tasks, from linear models to deep neural networks \cite{dean2012large,nedic2009distributed}.
	
	Stochastic gradient descent (SGD) is a widely used optimization algorithm for solving problem \eqref{eqn_1}. At each iteration $t$, the algorithm updates the parameter vector $\omega$ as
	\begin{equation}
		\label{eqn_2}
		\omega^{t+1}=\omega^{t}-\gamma \cdot \frac{1}{M} \sum_{m \in \mathcal{M}} \nabla f_{m}(\omega^{t}; \xi_{m}^{t}),
	\end{equation}
	where $\xi_{m}^{t}$ is a mini-batch data selected by worker $m$ at the $t$-th iteration, and $\gamma$ is the learning rate. In the context of distributed ML, parameter server (PS) is a common architecture that has been extensively studied in \cite{li2014scaling,li2014communication}. This architecture involves $M$ workers that parallelly compute the gradients based on their local data while a centralized server updates the global parameter $\omega$. The distributed SGD in PS operates as follows. At iteration $t$, the server broadcasts the current model $\omega^{t}$ to all workers, and each worker $m \in \mathcal{M}$ computes the local gradient $\nabla f_{m}(\omega^{t}; \xi_{m}^{t})$ with the mini-batch samples $\xi_{m}^{t} \sim \mathbb{D}_{m}$, which is then uploaded to the server. The server aggregates the received gradients and updates $\omega^{t}$ accordingly via \eqref{eqn_2}. It is worth noting that the convergence of this distributed SGD algorithm has been proved under mild assumptions \cite{zinkevich2010parallelized}.
	
	Intuitively, multi-processor collaborative training for one task can accelerate the training process and reduce training time. However, the associated communication costs typically hinder the scalability of distributed systems \cite{jordan2019communication}. More specifically, at each iteration of PS training \eqref{eqn_2}, the server needs to communicate with all workers to obtain fresh gradients $\{\nabla f_{m}(\omega^{t}; \xi_{m}^{t})\}_{m=1}^{M}$, which may lead to unaffordable communication overhead in distributed systems. Even worse, when the computation-to-communication ratio is low, e.g., using high-speed computing devices with low-speed interconnects to train a ML model, parallelization across multiple processors can actually result in lower performance than training on a single processor \cite{lin2018deep}. Consequently, the communication overhead between distributed workers and the server becomes a significant bottleneck.

	Various techniques have been proposed to alleviate this issue, including sparse communication \cite{aji2017sparse,NEURIPS2018_31445061,ElibolLJ20,stich2018sparsified} and adaptive aggregation methods \cite{chen2018lag,chen2021cada,chen2020lasg}. The former is dedicated to compressing each gradient information $\nabla f_{m}(\omega^{t}; \xi_{m}^{t})$, while the latter focuses on reducing the number of communications $M$. Since these two techniques are orthogonal, an important question arises: \emph{can we integrate both approaches to obtain a superior mechanism for problem \eqref{eqn_1}?}
	\IEEEpubidadjcol
	
	This study provides a positive answer. Specifically, we propose a communication-efficient distributed algorithm that simultaneously reduces the number of communication rounds and bits without compromising the convergence rate.
	
	\smallskip
	\noindent{\textbf{Contributions.}}
	This paper focuses on reducing the worker-to-server uplink communication overhead in the PS architecture, which we also refer to as upload. Unlike the server-to-worker downlink communication (i.e., broadcast the same parameter $\omega$), which can be performed concurrently or implemented in a tree-structured manner as in many MPI implementations, the server has to receive gradients of workers sequentially to avoid interference from other workers, resulting in extra latency. Consequently, uploads constitute the primary communication overhead in PS and are the subject of numerous related works \cite{aji2017sparse,alistarh2017qsgd,nips/0001DKD19,wen2017terngrad,chen2018lag,sun2019communication}. Throughout this paper, one communication round signifies one upload from a worker. The contributions of the paper are summarized below.
	\begin{itemize}
		\item We propose a communication-efficient algorithm for distributed learning that can reduce both the number of communication rounds and the number of bits.
		\item We introduce an auxiliary sequence and define a new Lyapunov function to establish the convergence analysis of the proposed algorithm. The theoretical results are consistent with the sublinear convergence rate of SGD and exhibit scalability to the number of distributed workers.
		\item We conduct extensive experiments that demonstrate the superiority of the proposed SASG algorithm.\footnote{The code is available at \url{https://github.com/xiaogdeng/SASG}}
	\end{itemize}
	
	\noindent{\textbf{Notation.}}
	In this paper, $\mathbb{E}[\cdot]$ represents the expectation for randomly selected data samples. For a vector $\mathbf{x} \in \mathbb{R}^{d}$ and a scalar $a \in \mathbb{R}$, $\|\mathbf{x}\|$ denotes the $\ell_{2}$-norm and $|a|$ denotes the absolute value. The notation $[d]$ represents the set $\{1, 2, \cdots, d\}$, and $|\mathcal{S}|$ indicates the number of elements in the set $\mathcal{S}$.

	\section{Related work}
	\label{sec:related}
	Numerous communication-efficient distributed learning approaches have been proposed to fully exploit the computational power of distributed clusters \cite{alistarh2017qsgd,nips/0001DKD19,chen2018lag,9515790,Samuel2021,stich2019local,NEURIPS2019_d9fbed9d,wang2018atomo,wen2017terngrad,li2021communication}. More information can be found in a comprehensive survey \cite{tang2020communication}. This paper briefly reviews two kinds of related work: 1). from the perspective of what to communicate, the number of transmitted bits per communication round can be reduced via quantization or sparsification, and 2). from the perspective of when to communicate, some communication rules are used to save the number of communication rounds.
	
	\subsection{Communication bit reduction}
	This research category mainly revolves around quantization and sparsification. The quantization approach compresses information by transmitting lower bits instead of the data originally represented by $32$ bits on each dimension of the transmitted gradient. Quantized stochastic gradient descent (QSGD) \cite{alistarh2017qsgd} utilizes an adjustable quantization level, allowing additional flexibility to control the trade-off between the per-iteration communication cost and the convergence rate. TernGrad \cite{wen2017terngrad} reduces the communication data size by using ternary gradients. 1-bit quantization method was developed in \cite{bernstein2018signsgd,seide20141,fan20221}, which reduces each component of the gradient to just its sign (one bit). Adaptive quantization methods are also investigated in \cite{Faghri2020agq} to reduce the communication cost.

	Sparsification methods aim to reduce the number of transmitted elements at each iteration. These methods can be divided into two main categories: random and deterministic sparsification. Random sparsification selects some entries randomly for communication \cite{wang2018atomo,wangni2018gradient}. This ideology is named random-$k$, with $k$ being the number of selected elements. This random choice method is usually an unbiased estimate of the original gradient, making it quite friendly for theoretical analysis. Unlike random sparsification, deterministic sparsification preserves only a few coordinates of the stochastic gradient with the largest magnitudes \cite{aji2017sparse,NEURIPS2018_31445061,ElibolLJ20,stich2018sparsified}. This ideology is also known as top-$k$. In contrast to the unbiased scheme, it is clear that this approach requires error feedback or accumulation procedures to ensure convergence \cite{karimireddy2019error,stich2018sparsified,xie2020cser}.

	\subsection{Communication round reduction}
	Reducing the number of communication rounds to improve communication efficiency is another research focus. Shamir et al. \cite{shamir2014communication} leveraged higher-order information (newton-type method) instead of traditional gradient information, thereby reducing the number of communication rounds. Hendrikx et al. \cite{Hendrikx2020spag} proposed a distributed preconditioned accelerated gradient method to reduce the number of communication rounds. Additionally, novel aggregation techniques, including periodic aggregation \cite{nips/0001DKD19,lin2019don,stich2019local} and adaptive aggregation \cite{chen2018lag,chen2021cada,sun2019communication}, have been explored to skip certain communications. Among these techniques, local SGD allows every worker to perform local model updates independently \cite{lin2019don,stich2019local}, and the resultant models are averaged periodically. Lazily aggregated gradient (LAG) \cite{chen2018lag} updates the model on the server side, and workers only adaptively upload information that is determined to be informative enough. Unfortunately, while the original LAG has good performance in the deterministic settings (i.e., with full gradient), its performance in the stochastic setting degrades significantly \cite{chen2020lasg}. Recent efforts have been made toward adaptive uploading in stochastic settings \cite{chen2020lasg,li2019communication}. Communication-censored distributed stochastic gradient descent (CSGD) algorithm \cite{li2019communication} increases the batch size to alleviate the effect of stochastic gradient noise. Lazily aggregated stochastic gradient (LASG) \cite{chen2020lasg} designed a set of new adaptive communication rules tailored for stochastic gradients, achieving remarkable empirical performance.

	We have drawn inspiration from LASG. The significant difference is that LASG only reduces communication round overhead, while our SASG method extends this idea to reduce the overhead of both communication rounds and bits in the stochastic setting. Additionally, SASG reduces the memory overhead required for adaptive aggregation technique by storing only the sparsified data on the server side.

	\begin{figure*}[t]
		\centering
		\includegraphics[width=0.75\textwidth]{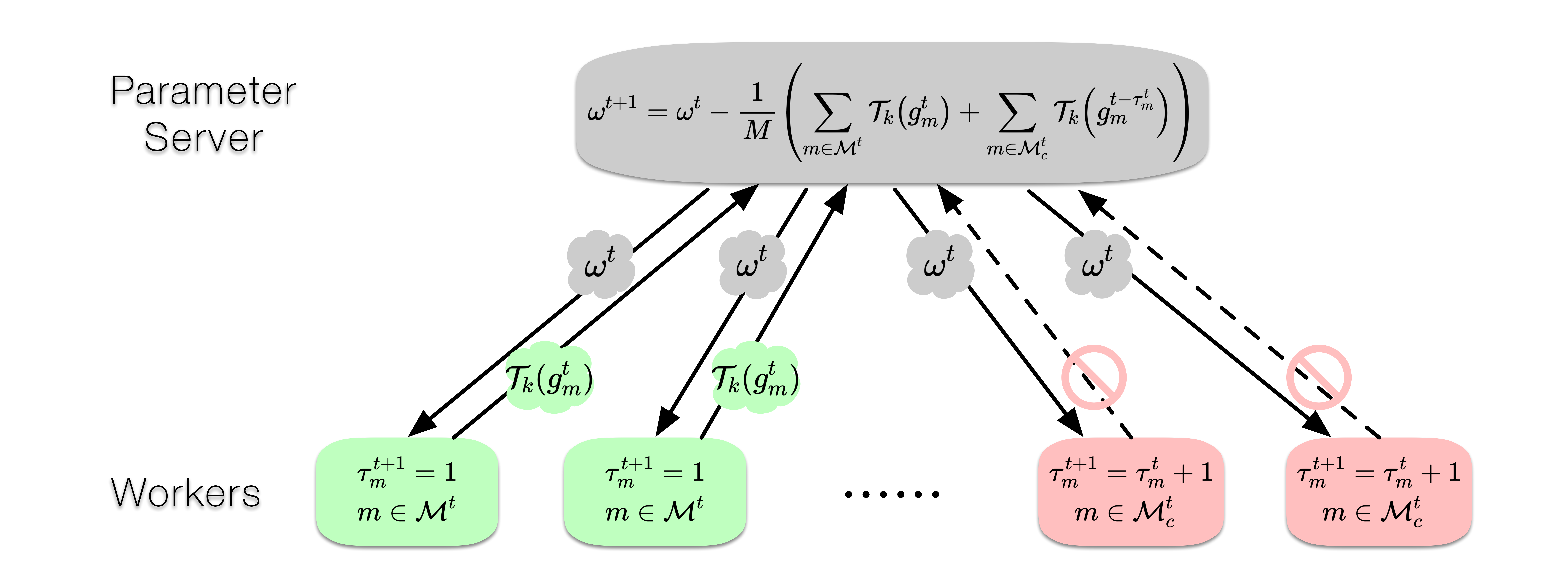}
		\caption{SASG overview. At the $t$-th iteration, the parameter server broadcasts ${\omega}^{t}$ to all workers, and workers in $\mathcal{M}^{t}$ (marked in green) will upload the sparsified gradient information $\mathcal{T}_{k}(g_{m}^{t})$ and reset $\tau_{m}^{t+1}=1$; while workers in $\mathcal{M}_{c}^{t}$ (marked in pink) will increase the staleness by $\tau_{m}^{t+1}=\tau_{m}^{t}+1$ and upload nothing; then the server updates parameter $\omega$ via \eqref{scheme}.}
		\label{fig:sasg}
	\end{figure*}

	\section{Algorithm development}
	\label{sec:algo}
	This section introduces SASG, a communication-efficient distributed algorithm that reduces both the number of communication rounds and bits. Firstly, we describe the motivation and challenges for developing this method, followed by the implementation details of the proposed algorithm.

	\subsection{Motivation and challenges}
	In the distributed learning system, an important observation is that not all communication rounds between the server and workers contribute equally to the state of the global model \cite{chen2018lag,chen2021cada,chen2020lasg,DBLP:conf/icdcs/WangWL19}. Then, one can design an aggregation rule to skip inefficient communication rounds. LAG has developed an adaptive selection rule that can detect workers with slowly varying gradients and trigger the reuse of outdated gradients to improve communication efficiency. However, this technique only reduces the number of communication rounds and requires additional memory overhead on the server side, which is far from enough in resource-limited scenarios. Actually, we can do more beyond LAG, e.g., further reduce the transmitted bits of uploaded information rather than just the number of rounds, and reduce the storage overhead on the server side.
	
	Two approaches for reducing the number of transmission bits are quantization and sparsification. We employ the latter as quantization methods are limited to a compression rate of up to $32\mathsf{x}$ (1-bit quantization method) in commonly used single-precision floating-point algorithms. When the model dimension $d$ is large, i.e., $d \gg 32$, the sparsification method will far outperform the quantization approach, offering a maximum achievable compression rate of $d\times$.
	
	Although random sparsification methods ensure the unbiasedness of the compression operator and facilitate theoretical analysis, the top-$k$ sparsification method tends to achieve better practical results \cite{wangni2018gradient,stich2018sparsified}. We will utilize the top-$k$ sparsification operator, denoted by $\mathcal{T}_{k}(\cdot)$. This operator retains only the largest $k$ components (in absolute value) of the gradient vector, and the specific definition is as follows.
	\begin{definition}
		\label{def:tk}
		For a parameter $1\leq k < d$, the sparsification operator $\mathcal{T}_{k}(\cdot): \mathbb{R}^{d}\rightarrow \mathbb{R}^{d}$, is defined for $\mathbf{x} \in \mathbb{R}^{d}$ as
		\begin{equation}
			\label{topk}
			\left(\mathcal{T}_{k}(\mathbf{x})\right)_{\pi(i)}:=
			\left\{\begin{array}{ll}
				(\mathbf{x})_{\pi(i)}, & \text { if } i \leq k, \\
				0, & \text { otherwise },
			\end{array}\right.
		\end{equation}
		where $\pi$ is a permutation of set $[d]$ such that $|(\mathbf{x})_{\pi(i)}| \geq|(\mathbf{x})_{\pi(i+1)}|$ for $i=1, \ldots, d-1$.
	\end{definition}
	
	Nevertheless, these motivations encounter two significant challenges: 1). How to design an effective communication criterion in the sparsification setting? That is, in this composite situation, we need to determine which communications are non-essential exactly. 2). How can we guarantee model convergence when updating the parameters with only sparsified information from a subset of workers? In particular, the only partial gradient information we have is still biased. Our proposed approach will address these challenges.

	\subsection{SASG method}
	To tackle the first challenge, we need to determine an effective selection criterion. Intuitively, if the difference between two consecutive gradients from a worker is small, it is safe to skip the redundant uploads and reuse the previous one stored on the server. The difference is defined as
	\begin{equation}
		\label{delta}
		\Delta_{m}^{t} := \nabla f_{m}({\omega}^{t}; \xi_{m}^{t})-\nabla f_{m}({\omega}^{t-\tau_{m}^{t}}; \xi_{m}^{t-\tau_{m}^{t}}),
	\end{equation}
	where $\tau_{m}^{t}$ is the delay count. Additionally, we need to pick a threshold that varies with iteration $t$ to measure the magnitude of the difference $\Delta_{m}^{t}$ adaptively. Following the idea of LAG, we have
	\begin{equation}
		\label{lag}
		\Delta_{m}^{t} \leq \frac{1}{M^{2}} \sum_{d=1}^{D}\alpha_{d}\left\|\omega^{t+1-d}-\omega^{t-d}\right\|^{2},
	\end{equation}
	where $\{\alpha_{d}\geq 0\}_{d=1}^{D}$ are constant weights. This rule is a direct extension of LAG to the stochastic settings. Unfortunately, \eqref{lag} is ineffective due to the non-diminishing variance of the stochastic gradients \eqref{delta}. Specifically, the variance introduced by the randomly selected samples $\xi_{m}^{t}$ and $\xi_{m}^{t-\tau_{m}^{t}}$ across two iterations makes the value of $\Delta_{m}^{t}$ almost never small, rendering criterion \eqref{lag} less effective. Our SASG method takes advantage of the adaptive selection rule from LASG \cite{chen2020lasg}, formulated as
	\begin{equation}
		\label{condition}
		\begin{split}
			&\left\|\nabla f_{m}({\omega}^{t}; \xi_{m}^{t})-\nabla f_{m}({\omega}^{t-\tau_{m}^{t}}; \xi_{m}^{t})\right\|^{2} \\
			&\leq  \sum_{d=1}^{D}\frac{\alpha_{d}}{M^{2}}\left\|\omega^{t+1-d}-\omega^{t-d}\right\|^{2}.
		\end{split}
	\end{equation}

	This condition is evaluated on the same data $\xi_{m}^{t}$ at two different iterations $t$ and $t-\tau_{m}^{t}$. In the following, we provide a novel insight to demonstrate the suitability of the criterion for our method. According to the Lipschitz continuous property of the function $\nabla f_{m}$, we have
	\begin{equation}
		\label{lipachitz}
		\begin{split}
			\left\|\nabla f_{m}({\omega}^{t}; \xi_{m}^{t})\!-\!\nabla f_{m}({\omega}^{t-\tau_{m}^{t}}; \xi_{m}^{t})\right\|^{2}\!\leq\! L_{m}\left\|\omega^{t}\!-\!\omega^{t-\tau_{m}^{t}}\right\|^{2},
		\end{split}
	\end{equation}
	where $L_{m}$ is the Lipschitz constant. As the iterative sequence $\{\omega^{t}\}_{t=0,1,\ldots}$ converges, the right-hand side of \eqref{lipachitz} diminishes, and thus the left-hand side of \eqref{condition} diminishes, which eliminates the inherent variance caused by stochastic data $\xi_{m}^{t}$ and $\xi_{m}^{t-\tau_{m}^{t}}$. Additionally, this insight contributes to the determination of the threshold hyper-parameters in criterion \eqref{condition}.
	
	At each training iteration $t$, selection rule \eqref{condition} divides the worker set $\mathcal{M}$ into two disjoint sets, $\mathcal{M}^{t}$ and $\mathcal{M}_{c}^{t}$. The parameter server only needs to receive the new gradient data from $\mathcal{M}^{t}$ and reuse outdated gradients (stored on the server side) from $\mathcal{M}_{c}^{t}$, which scale down the per-iteration communication rounds from $M$ to $|\mathcal{M}^{t}|$. After the adaptive selection procedure, selected workers transmit the sparse information derived by the top-$k$ operator to the parameter server, which reduces the communication bits and the memory overhead on the server side. However, the biased nature of the top-$k$ operator disrupts algorithmic convergence, intensifying challenge two.

	\begin{algorithm}[t]
		\caption{SASG algorithm.}
		\label{alg:sasg}
		\setstretch{1.1}
		\mbox{\textbf{Initialize}: Errors ${e}_{m}^{0}=0$, delay counters $\tau_{m}^{0}=1$, $\forall m \in \mathcal{M}$}\\
		\textbf{Input}: Learning rate $\gamma>0$, maximum delay $D$, Constant weights $\{\alpha_{d}\}_{d=1}^{D}$
		\begin{algorithmic}[1] 
			\FOR{$t=0, 1, \ldots, T$}
			\STATE Server broadcasts ${\omega}^{t}$ to all workers.
			\FOR {worker $m=1, \ldots, M$}
			\STATE Compute $g_{m}^{t}=\gamma\nabla f_{m}({\omega}^{t}; \xi_{m}^{t})+e_{m}^{t}$.
			\STATE Divide the workers into $\mathcal{M}^{t}$ and $\mathcal{M}_{c}^{t}$ according to the selection rule \eqref{condition}.
			\IF {worker $m\in \mathcal{M}^{t}$ or $\tau_{m}^{t} \geq D$}
			\STATE Worker $m$ uploads $\mathcal{T}_{k}({g}_{m}^{t})$ to the server.
			\STATE Set ${e}_{m}^{t+1}={g}_{m}^{t}-\mathcal{T}_{k}({g}_{m}^{t}), ~~ \tau_{m}^{t+1}=1$.
			\ELSE
			\STATE Worker $m$ uploads nothing.
			\STATE Set ${e}_{m}^{t+1}={e}_{m}^{t}, ~~ \tau_{m}^{t+1}=\tau_{m}^{t}+1$.
			\ENDIF	
			\ENDFOR
			\STATE Server updates parameter ${\omega}$ according to \eqref{scheme}.
			\ENDFOR
		\end{algorithmic}
	\end{algorithm}
	To address this issue, SASG incorporates error feedback techniques, also known as error accumulation or memory \cite{karimireddy2019error,stich2018sparsified}. Initially, the algorithm sparsifies the gradient information $g$ to obtain $\mathcal{T}_{k}(g)$. When uploading the sparsified information $\mathcal{T}_{k}(g)$, we calculate the compression error $e\!=\!g\!-\!\mathcal{T}_{k}(g)$ and retain it for subsequent use. The next iteration involves adding the saved error $e$ to the gradient information and compressing them jointly. Eventually, all the gradient information will be transmitted despite the delay caused by the error feedback mechanism. We introduce an auxiliary sequence $\{\nu^{t}\}_{t=0,1,\ldots}$, which can be regarded as an error approximation of $\{\omega^{t}\}_{t=0,1,\ldots}$. Leveraging the recursive properties of this sequence, we prove that SASG is convergent in Section \ref{sec:analysis}.

	In summary, the iterative scheme of the SASG algorithm can be expressed as 
	\begin{equation}
		\setlength{\belowdisplayskip}{1pt}
		\label{scheme}
		\omega^{t+1}=\omega^{t}-\frac{1}{M}\bigg[\sum_{m \in \mathcal{M}^{t}}\mathcal{T}_{k}(g_{m}^{t})+\sum_{m \in \mathcal{M}_{c}^{t}}\mathcal{T}_{k}(g_{m}^{t-\tau_{m}^{t}})\bigg],
	\end{equation}
	where
	\begin{equation}
		g_{m}^{t}:=\gamma\nabla f_{m}({\omega}^{t}; \xi_{m}^{t})+e_{m}^{t},\quad {e}_{m}^{t+1}:={g}_{m}^{t}-\mathcal{T}_{k}({g}_{m}^{t}).
	\end{equation}
	Here, $\mathcal{M}^{t}$ and $\mathcal{M}_{c}^{t}$ denote the sets of workers that do and do not communicate with the server at the $t$-th iteration, respectively. Staleness $\tau_{m}^{t}$ is determined by the selection of the subset $\mathcal{M}^{t}$ at the $t$-th iteration. For worker $m \in \mathcal{M}^{t}$, the server resets $\tau_{m}^{t+1}=1$, and the worker uploads the local gradient information. Otherwise, the server increases staleness by $\tau_{m}^{t+1}=\tau_{m}^{t}+1$, and worker $m$ uploads nothing. The SASG procedure is illustrated in Figure \ref{fig:sasg} and outlined in Algorithm \ref{alg:sasg}. Specifically, during each iteration $t=0,1,2,\ldots$,
	\begin{enumerate}[i.]
		\item server broadcasts the learning parameter $\omega^{t}$ to all workers;
		\item each worker $m$ calculates the local gradient $\nabla f_{m}({\omega}^{t}; \xi_{m}^{t})$ and an auxiliary gradient $\nabla f_{m}({\omega}^{t-\tau_{m}^{t}}; \xi_{m}^{t})$ for computing the selection rule \eqref{condition};
		\item workers in $\mathcal{M}^{t}$ identified by condition \eqref{condition} sparsify the local information $g_{m}^{t}=\gamma\nabla f_{m}({\omega}^{t}; \xi_{m}^{t})+e_{m}^{t}$, and upload $\mathcal{T}_{k}({g}_{m}^{t})$ to the server;
		\item server aggregates the fresh sparsified gradients $\mathcal{T}_{k}(g_{m}^{t})$ from the selected workers $\mathcal{M}^{t}$ and the outdated gradient information $\mathcal{T}_{k}(g_{m}^{t-\tau_{m}^{t}})$ (stored on the server) from $\mathcal{M}_{c}^{t}$ to update the parameter $\omega$ via \eqref{scheme}.
	\end{enumerate}
	
	As shown in Table \ref{tab:overhead}, we only need to upload the sparse gradient information from the selected workers, thus significantly saving the communication overhead (note that $\mathcal{M}^{t}$ selected in SASG and LASG is not the same due to different update information). Furthermore, SASG stores the sparsified gradient on the server, which reduces the memory overhead of traditional adaptive aggregation methods.

	\begin{table*}[t]
		\begin{center}
			{\caption{Comparison of different algorithms, including the number of communication rounds per iteration, communication bits per upload, and the total communication overhead to complete $T$ iterations for each algorithm when training the $d$-dimensional parametric model with $M$ distributed workers ($k$ is the sparsification level defined in \eqref{topk}, which satisfies $k<d$).}
			\label{tab:overhead}}
			\begin{tabular}{ccccccc}
				\toprule[1pt]
				\rule{0pt}{10pt}
				Method & \# Round & \# Bit & Total Overhead & Skip communication? & Compress information? & Biased estimation?\\
				\midrule[0.5pt]
				\rule{0pt}{12pt}
				SGD & $M$ & $32d$ & $32dMT$ & \ding{56}  & \ding{56} & \ding{56}\\
				\rule{0pt}{15pt}
				Sparse & $M$ & $32k$ & $32kMT$ & \ding{56}  & \ding{52} & \ding{52}\\
				\rule{0pt}{15pt}
				LASG & $|\mathcal{M}^{t}|$ & $32d$ & $32d \cdot \sum\limits_{t=1}^{T}|\mathcal{M}^{t}|$ & \ding{52} & \ding{56} & \ding{56} \\
				\rule{0pt}{15pt}
				SASG & $|\mathcal{M}^{t}|$ & $32k$ & $32k \cdot \sum\limits_{t=1}^{T}|\mathcal{M}^{t}|$ & \ding{52} & \ding{52} & \ding{52} \\
				\bottomrule[1pt]
			\end{tabular}
		\end{center}
	\end{table*}

	\section{Theoretical analysis}
	\label{sec:analysis}
	
	This section provides a detailed theoretical analysis of the SASG algorithm. As shown in Table \ref{tab:overhead}, unlike LASG, the SASG algorithm relies solely on biased gradient information from partial workers for model updates, which complicates its convergence analysis. To overcome these challenges, we first prove that the residuals of the compression operator are bounded (Lemmas \ref{lem:1} and \ref{lem:2}) and introduce a critical auxiliary variable that couples the model parameters with the compression error. Subsequently, we define a new Lyapunov function with respect to the auxiliary variable, which is also compatible with the selection criterion \eqref{condition}. The descent lemmas on the value of the objective function (Lemma \ref{lem:3}) and the Lyapunov function (Lemma \ref{lem:4}) derive the convergence theorem for the SASG algorithm (Theorem \ref{thm:1}), and its convergence rate is consistent with that of the original SGD algorithm. All proof details are included in the Proofs (Section \ref{sec:proof}).

	\subsection{Assumptions}
	Our analysis is based on the following assumptions, which are standard in analyzing SGD and its variants. The functions $f_m, F_m$, and $F$ have been specified in Section \ref{sec:intro}.
	\begin{assumption}[Smoothness] 
		\label{amp:1}
		The loss function $f_{m}: \mathbb{R}^{d}\rightarrow \mathbb{R}$ is $L_{m}$-smooth and function $F: \mathbb{R}^{d}\rightarrow \mathbb{R}$ is $L$-smooth, i.e., $\forall \mathbf{x}, \mathbf{y} \in \mathbb{R}^{d}$,
		\begin{equation}
			\begin{split}
				\|\nabla f_{m}(\mathbf{y})-\nabla f_{m}(\mathbf{x})\| &\leq L_{m}\|\mathbf{y}-\mathbf{x}\|,\\
				\|\nabla F(\mathbf{y})-\nabla F(\mathbf{x})\| &\leq L\|\mathbf{y}-\mathbf{x}\|.
			\end{split}
		\end{equation}
	\end{assumption}
	
	\begin{assumption}[Moment boundedness]
		\label{amp:2}
		The data sampled in each iteration, i.e., $\xi_{m}^{1}, \xi_{m}^{2}, \ldots$, are independent, and the stochastic gradient $\nabla f_{m}(\omega; \xi_{m}^{t})$ satisfies
		\begin{equation}
			\begin{array}{l}
				\mathbb{E} [\nabla f_{m}(\omega; \xi_{m}^{t})]=\nabla F_{m}(\omega),\\
				\mathbb{E}\left [\|\nabla f_{m}(\omega; \xi_{m}^{t})-\nabla F_{m}(\omega)\|^{2}\right] \leq \sigma_{m}^{2}.
			\end{array}
		\end{equation}
	\end{assumption}
	
	\begin{assumption}[Gradient boundedness]
		\label{amp:3}
		For a given parameter $\omega \in \mathbb{R}^{d}$, the local gradient $\nabla f_{m}(\omega; \cdot)$ is bounded, i.e., there exists a constant $B_{m}\in \mathbb{R}$ such that
		\begin{equation}
			\mathbb{E}\left[\|\nabla f_{m}(\omega; \cdot)\|\right] \le B_{m}.
		\end{equation}
	\end{assumption}

	\subsection{Technical lemmas}
	The analysis commences by highlighting a key property of the top-$k$ sparsification operator $\mathcal{T}_{k}(\cdot)$.
	\begin{lemma}
		\label{lem:1}
		The operator $\mathcal{T}_{k}(\cdot)$ is a $\delta$-approximate compressor, i.e., there exists a constant $\delta=k/d \in (0, 1)$ such that
		\begin{equation}
			\|\mathcal{T}_{k}(\mathbf{x})-\mathbf{x}\|^{2} \leq(1-\delta)\|\mathbf{x}\|^{2}, \forall \mathbf{x} \in \mathbb{R}^{d}.
		\end{equation}
	\end{lemma}
	With the gradient boundedness assumption, we are prepared to present a critical lemma, which demonstrates how to bound the residual error in Algorithm \ref{alg:sasg}.
	
	\begin{lemma}
		\label{lem:2}
		Under Assumption \ref{amp:3}, at each iteration $t$ of the SASG algorithm, the residual error $e_{m}^{t}$ in worker $m$ is bounded, i.e., let $\delta=k/d$, we have
		\begin{equation}
			\mathbb{E}\left[\|e_{m}^{t}\|^{2}\right] \leq \frac{4(1-\delta)}{\delta^{2}}\gamma^{2}B_{m}^{2}.
		\end{equation}
	\end{lemma}
	Lemma \ref{lem:2} demonstrates that the residual errors maintained in Algorithm \ref{alg:sasg} do not accumulate too much and are closely related to the sparsification coefficient $k$. Next, we introduce a crucial auxiliary sequence for subsequent analysis.

	\begin{definition}
		\label{def:v}
		Let $\{\omega^{t}\}_{t=0,1,\ldots}$ be the sequence generated by SASG, and the auxiliary variable $\nu^{t}$ is defined as
		\begin{equation}
			\label{v:def}
			\nu^{t}:=\omega^{t}-\frac{1}{M} \sum_{m\in \mathcal{M}}e_{m}^{t}.
		\end{equation}
	\end{definition}
	The sequence $\{\nu^{t}\}_{t=0,1,\ldots}$, which can be considered as an error-corrected sequence of $\{\omega^{t}\}_{t=0,1,\ldots}$, has the following property ($\Delta_{m}^{t}$ is defined in \eqref{delta})
	\begin{equation}
		\label{aux:v}
		\nu^{t+1}-\nu^{t}=-\frac{\gamma}{M} \sum_{m \in \mathcal{M}}\nabla f_{m}({\omega}^{t}; \xi_{m}^{t})+\frac{\gamma}{M} \sum_{m \in \mathcal{M}_{c}^{t}}\Delta_{m}^{t},
	\end{equation}
	Based on this property, we have the following descent lemma for the loss function concerning the auxiliary variable $\nu^{t}$.
	
	\begin{lemma}
		\label{lem:3}
		Let Assumptions \ref{amp:1} and \ref{amp:2} hold, and the sequence $\{\nu^{t}\}_{t=0,1,\ldots}$ is defined in \eqref{v:def}. Then, the objective function value satisfies (with $\sigma^{2}:=\sum_{m=1}^{M}\sigma_{m}^{2}$)
		\begin{equation}
			\label{lemma3}
			\begin{split}
				&\mathbb{E}[F({\nu}^{t+1})]-\mathbb{E}[F({\nu}^{t})] \leq -(\gamma-\frac{5L\!+\!4}{2}\gamma^2)\mathbb{E}\left[\|\nabla F({\omega}^{t})\|^{2}\right]\\
				&~~+\sum_{d=1}^{D}\bigg[\frac{\alpha_{d}}{2M^{2}L}+\frac{L}{2}+\frac{2\gamma^2(L\!+\!1)\alpha_{d}}{M^{2}}\bigg]\|\omega^{t+1-d}-\omega^{t-d}\|^{2}\\
				&~~+\frac{L^{2}}{2M^{2}}\mathbb{E}\bigg[\Big\|\sum_{m \in \mathcal{M}}e_{m}^{t}\Big\|^{2}\bigg]+\left[10(L+1)+2DL\right]\frac{\gamma^{2}\sigma^{2}}{M}.
			\end{split}
		\end{equation}
	\end{lemma}
	
	It is worth noting that all terms on the right-hand side of inequality \eqref{lemma3} exist in SGD analysis except $\|\omega^{t+1-d}-\omega^{t-d}\|^{2}$ and $\|\sum_{m \in \mathcal{M}}e_{m}^{t}\|^{2}$, which arise due to stale information and compression errors. To deal with these terms, we will introduce an associated Lyapunov function. With $F^{*}$ denoting the optimal value of problem \eqref{eqn_1}, the Lyapunov function is defined as
	\begin{equation}
		\label{lya}
		\mathcal{L}^{t}:=\mathbb{E}[F(\nu^{t})]-F^{*}+\sum_{d=1}^{D}\beta_{d}\|\omega^{t+1-d}-\omega^{t-d}\|^{2},
	\end{equation}
	where $\{\beta_d\ge0\}_{d=1}^{D}$ are constants that will be determined later. The Lyapunov function is coupled with the selection rule \eqref{condition} that contains the parameter difference terms. We also highlight that in the definition \eqref{lya}, $\mathcal{L}^{t}>0$ for any $t \in \mathbb{N}$. A direct extension of Lemma \ref{lem:3} gives the following descent lemma.

	\begin{lemma}
		\label{lem:4}
		Let Assumptions \ref{amp:1}, \ref{amp:2}, and \ref{amp:3} hold and denote $\sigma^{2}:=\sum_{m=1}^{M}\sigma_{m}^{2}$, $B^{2}:=\sum_{m=1}^{M}B_{m}^{2}$. If the learning rate $\gamma$ and constant weights $\{\alpha_{d}\}_{d=1}^{D}$ are chosen properly, the Lyapunov function satisfies
		\begin{equation}
			\begin{split}
				\mathcal{L}^{t+1}-\mathcal{L}^{t} \le & -c_{f} \mathbb{E}[\|\nabla F({\omega}^{t})\|^{2}]+a\frac{\gamma^{2}\sigma^{2}}{M}+b\frac{\gamma^{2}B^{2}}{M}\\
				&-\sum_{d=1}^{D} c_{d}\|\omega^{t+1-d}-\omega^{t-d}\|^{2},
			\end{split}
		\end{equation}
		where $c_{f}, a, b, c_{1}, \ldots, c_{D} \ge 0$ depend on the learning rate $\gamma$, constants $D, L,$ and $\{\alpha_{d}, \beta_{d}\}_{d=1}^{D}$. More specific explanatory notes can be found in the proof details.
	\end{lemma}
	
	\subsection{Main results}
	In conjunction with the lemmas introduced above, we are ready to present the main convergence results of our SASG algorithm.
	\begin{Theorem}
		\label{thm:1}
		Under Assumptions \ref{amp:1}, \ref{amp:2}, and \ref{amp:3}, let the sequence $\{\omega^{t}\}_{t=0,1,\ldots}$ be generated by the SASG algorithm. If constant weights $\{\alpha_{d}\}_{d=1}^{D}$ are selected properly, and the learning rate is chosen as $\gamma=\min \{1/(5L+4+16\beta_{1}), c_{\gamma}/\sqrt{T}\}$, where $c_{\gamma}>0$ is a constant and $\beta_{1}$ is defined in \eqref{lya}, we then have
		\begin{equation}
			\label{eqn:thm1}
			\begin{split}
				&\frac{1}{T}\sum_{t=0}^{T-1}\mathbb{E}\left[\|\nabla F({\omega}^{t})\|^{2}\right] \leq \frac{2(5L+4+16\beta_{1})(F(\omega^{0})-F^{*})}{T} \\
				&\qquad\qquad\qquad\qquad+\frac{2(F(\omega^{0})-F^{*})}{c_{\gamma}\sqrt{T}} +\frac{2c_{\gamma}(a\sigma^{2}+bB^{2})}{M\sqrt{T}}.
			\end{split}
		\end{equation}
	\end{Theorem}

	Theorem \ref{thm:1} establishes the convergence guarantee of our algorithm with a sublinear convergence rate of $\mathcal{O}(1/\sqrt{T})$, despite skipping many communication rounds and performing communication compression. In other words, the SASG algorithm, utilizing well-designed adaptive aggregation and sparse communication techniques, can still achieve a convergence rate identical to that of the SGD method. 
	
	\begin{remark}
		In the analysis of Theorem \ref{thm:1}, we also delved into the effect of the number of workers $M$ on the convergence performance. As indicated by result \eqref{eqn:thm1}, the convergence is controlled by the average variance of the distributed system and is not affected by the number of workers, making SASG scale well to large-scale distributed systems.
	\end{remark}

	\begin{remark}
		It follows from the proof of the theorem that the constant $b$ is defined as
		\begin{equation}
			\nonumber
			b:=\frac{(2L^{2}+16\beta_{1})(1-\delta)}{\delta^{2}}, \quad \delta=\frac{k}{d}.
		\end{equation}
		As the sparsification operation retains more components (i.e., with a larger value of $k$), the parameter $b$ diminishes, and thus it is more favorable for the algorithm to converge. This theoretical finding is consistent with empirical intuition. On the other hand, it is important to recognize that the convergence result of the algorithm depends on more than the term $\frac{2c_{\gamma}bB^{2}}{M\sqrt{T}}$. Consequently, using different sparse coefficients cannot change the sublinear convergence rate of the algorithm.
	\end{remark}

	\section{Proofs}
	\label{sec:proof}
	This section contains proof details of Lemmas and Theorem in Section \ref{sec:analysis}, and all the theoretical proofs are based on the standard Assumptions \ref{amp:1}, \ref{amp:2}, and \ref{amp:3}.
	
	\subsection{Proof of Lemma \ref{lem:1}}
	\label{app:a-1}
	\noindent Given a vector $\mathbf{x} \in \mathbb{R}^{d}$, we observe from Definition \ref{def:tk} that only the first $k$ largest coordinates of the vector are retained. Then, we can derive
	\begin{equation*}
		\|\mathcal{T}_{k}(\mathbf{x})\|^{2} \geq \frac{k}{d}\|\mathbf{x}\|^{2}.
	\end{equation*}
	Since $\left\langle \mathcal{T}_{k}(\mathbf{x})-\mathbf{x}, \mathcal{T}_{k}(\mathbf{x})\right\rangle=0$, it follows that 
	\begin{equation*}
		\|\mathcal{T}_{k}(\mathbf{x})-\mathbf{x}\|^{2}=\|\mathbf{x}\|^{2}-\|\mathcal{T}_{k}(\mathbf{x})\|^{2}.
	\end{equation*}
	Let $\delta:=k/d \in (0, 1)$, we can deduce that
	\begin{equation*}
		\|\mathcal{T}_{k}(\mathbf{x})-\mathbf{x}\|^{2}\leq(1-\delta)\|\mathbf{x}\|^{2}.
	\end{equation*}
	
	\subsection{Proof of Lemma \ref{lem:2}}
	\label{app:a-2}
	\noindent From Algorithm \ref{alg:sasg} and Lemma \ref{lem:1}, it follows that
	\begin{equation*}
		\begin{split}
			&\mathbb{E}[\|{e}_{m}^{t+1}\|^{2}]=\mathbb{E}[\|{g}_{m}^{t}-\mathcal{T}_{k}({g}_{m}^{t})\|^{2}]\leq(1-\delta)\mathbb{E}[\|{g}_{m}^{t}\|^{2}]\\
			&=(1-\delta)\mathbb{E}\left[\|\gamma \nabla f_{m}({\omega}^{t}, \xi_{m}^{t})+{e}_{m}^{t}\|^{2}\right]\\
			&\le (1\!-\!\delta)\Big((1\!+\!\eta)\mathbb{E}\|{e}_{m}^{t}\|^{2}+(1\!+\!\frac{1}{\eta})\gamma^{2}\mathbb{E}\|\nabla f_{m}({\omega}^{t},\xi_{m}^{t})\|^{2}\Big),
		\end{split}
	\end{equation*}
	where we used Young's inequality $\|a+b\|^{2} \le (1+\eta)\|a\|^{2}+(1+1/\eta)\|b\|^{2}$ for any $\eta>0$. By Algorithm \ref{alg:sasg} and Assumption \ref{amp:3}, we have that error ${e}_{m}^{0}=0$ and gradient $\nabla f_{m}({\omega}^{t},\xi_{m}^{t})$ is bounded. Simple computation yields
	\begin{equation*}
		\begin{split}
			&\mathbb{E}[\|{e}_{m}^{t+1}\|^{2}]\\
			&\le \sum_{l=0}^{t}\big[(1\!-\!\delta)(1\!+\!\eta)\big]^{t-l}(1\!-\!\delta)(1\!+\!\frac{1}{\eta})\gamma^{2}\mathbb{E}\|\nabla f_{m}({\omega}^{l},\xi_{m}^{l})\|^{2}\\
			&=\frac{(1-\delta)(1+1 / \eta)}{1-(1-\delta)(1+\eta)}\gamma^{2} B_{m}^{2}.
		\end{split}
	\end{equation*}
	By selecting $\eta:=\frac{\delta}{2(1-\delta)}$ yielding $1+1/\eta=(2-\delta)/\delta \leq 2/\delta$, we arrive at
	\begin{equation*}
		\mathbb{E}[\|{e}_{m}^{t+1}\|^{2}]\leq\frac{2(1-\delta)(1+1 / \eta)}{\delta}\gamma^{2} B_{m}^{2} \leq \frac{4(1-\delta)}{\delta^{2}} \gamma^{2}B_{m}^{2}.
	\end{equation*}

	\subsection{Proof of Equation \eqref{aux:v}}
	\label{app:a-3}
	\noindent Following Definition \ref{def:v} and the iterative scheme \eqref{scheme}, we have
	\begin{equation*}
		\begin{split}
			&\nu^{t+1}=\omega^{t+1}-\frac{1}{M}\sum_{m\in \mathcal{M}}e_{m}^{t+1}\\
			&=\omega^{t}-\frac{1}{M}\bigg[\sum_{m \in \mathcal{M}^{t}}\mathcal{T}_{k}(g_{m}^{t})\!+\!\sum_{m \in \mathcal{M}_{c}^{t}}\mathcal{T}_{k}(g_{m}^{t-\tau_{m}^{t}})\!+\!\sum_{m\in \mathcal{M}}e_{m}^{t+1}\bigg]\\
			&\overset{*}{=}\omega^{t}-\frac{1}{M}\bigg[\sum_{m \in \mathcal{M}^{t}}g_{m}^{t}+\sum_{m \in \mathcal{M}_{c}^{t}}g_{m}^{t-\tau_{m}^{t}}\bigg]\\
			&=\omega^{t}-\frac{1}{M}\sum_{m \in \mathcal{M}^{t}}\left(\gamma \nabla f_{m}({\omega}^{t}; \xi_{m}^{t})+e_{m}^{t}\right)\\
			&\quad -\frac{1}{M}\sum_{m \in \mathcal{M}_{c}^{t}}\left(\gamma \nabla f_{m}({\omega}^{t-\tau_{m}^{t}}; \xi_{m}^{t-\tau_{m}^{t}})+e_{m}^{t-\tau_{m}^{t}}\right)\\
			&\overset{*}{=}\nu^{t}-\frac{\gamma}{M}\sum_{m\in \mathcal{M}}\nabla f_{m}({\omega}^{t}; \xi_{m}^{t})\\
			&\quad +\frac{\gamma}{M}\sum_{m \in \mathcal{M}_{c}^{t}}\left(\nabla f_{m}({\omega}^{t}; \xi_{m}^{t
			})-\nabla f_{m}({\omega}^{t-\tau_{m}^{t}}; \xi_{m}^{t-\tau_{m}^{t}})\right),
		\end{split}
	\end{equation*}
	where $(*)$ used $e_{m}^{t+1}=e_{m}^{t-\tau_{m}^{t}}, \forall m \in \mathcal{M}_{c}^{t}$. Let
	\begin{equation*}
	\Delta_{m}^{t} := \nabla f_{m}({\omega}^{t}; \xi_{m}^{t})-\nabla f_{m}({\omega}^{t-\tau_{m}^{t}}; \xi_{m}^{t-\tau_{m}^{t}}),
	\end{equation*}
	it then follows that
	\begin{equation}
	\label{aux:appv}
	\nu^{t+1}-\nu^{t}=-\frac{\gamma}{M} \sum_{m \in \mathcal{M}}\nabla f_{m}({\omega}^{t}; \xi_{m}^{t})+\frac{\gamma}{M} \sum_{m \in \mathcal{M}_{c}^{t}}\Delta_{m}^{t}.
	\end{equation}

	\subsection{Proof of Lemma \ref{lem:3}}
	\label{app:a-4}
	\noindent We first present the complete derivation and then delve into the corresponding specifics.
	\begin{equation}
	\label{F_descent}
	\begin{split}
		&\mathbb{E}\left[F(\nu^{t+1})\right]-\mathbb{E}\left[F(\nu^{t})\right]\\
		&\overset{(a)}{\leq} \mathbb{E}\Big\langle\nabla F(\nu^{t}), \nu^{t+1}-\nu^{t}\Big\rangle+\frac{L}{2}\mathbb{E}\|\nu^{t+1}-\nu^{t}\|^{2}\\
		&\overset{(b)}{\leq} \mathbb{E}\Big\langle \nabla F({\omega}^{t}), -\frac{\gamma}{M}\sum_{m \in \mathcal{M}}\nabla f_{m}({\omega}^{t}; \xi_{m}^{t})+\frac{\gamma}{M}\sum_{m \in \mathcal{M}_{c}^{t}}\Delta_{m}^{t}\Big\rangle \\
		&\quad +\mathbb{E}\left\langle\nabla F(\nu^{t})-\nabla F(\omega^{t}), \nu^{t+1}-\nu^{t}\right\rangle+\frac{L}{2}\mathbb{E}\|\nu^{t+1}-\nu^{t}\|^{2}\\
		&\overset{(c)}\leq -\gamma\mathbb{E}\|\nabla F({\omega}^{t})\|^{2}+\frac{\gamma}{M} \mathbb{E}\Big\langle\nabla F({\omega}^{t}), \sum_{m \in \mathcal{M}_{c}^{t}}\Delta_{m}^{t}\Big\rangle \\
		&\quad +\frac{1}{2}\mathbb{E}\|\nabla F(\nu^{t})-\nabla F({\omega}^{t})\|^{2}+\frac{L+1}{2}\mathbb{E}\|\nu^{t+1}-\nu^{t}\|^{2}\\
		&\overset{(d)}\leq -(\gamma-\frac{L\gamma^2}{2})\mathbb{E}\|\nabla F({\omega}^{t})\|^{2}+\frac{L+1}{2}\mathbb{E}\|\nu^{t+1}-\nu^{t}\|^{2}\\
		&\quad +\sum_{d=1}^{D}\left[\frac{\alpha_{d}}{2M^{2}L}+\frac{L}{2}\right]\|\omega^{t+1-d}-\omega^{t-d}\|^{2}\\
		&\quad +\frac{L^{2}}{2}\mathbb{E}\left[\|\nu^{t}-\omega^{t}\|^{2}\right]+\frac{2\gamma^{2}DL\sigma^{2}}{M}\\
		&\overset{(e)}\leq -(\gamma-\frac{5L+4}{2}\gamma^2)\mathbb{E}\left[\|\nabla F({\omega}^{t})\|^{2}\right]\\
		&\quad +\sum_{d=1}^{D}\left[\frac{\alpha_{d}}{2M^{2}L}+\frac{L}{2}+\frac{2\gamma^2(L\!+\!1)\alpha_{d}}{M^{2}}\right]\|\omega^{t+1-d}\!-\!\omega^{t-d}\|^{2}\\
		&\quad +\frac{L^{2}}{2M^{2}}\mathbb{E}\bigg[\Big\|\sum_{m \in \mathcal{M}}e_{m}^{t}\Big\|^{2}\bigg]+\left[10(L+1)+2DL\right]\frac{\gamma^{2}\sigma^{2}}{M},
	\end{split}
	\end{equation}
	where $(a)$ uses the $L$-smooth property outlined in Assumption \ref{amp:1}. $(b)$ is obtained by equation \eqref{aux:appv}. $(c)$ utilizes Assumption \ref{amp:2} and the inequality $\|a\|^{2}+\|b\|^{2}\geq 2\left\langle a, b\right\rangle$. To establish $(d)$, we need to bound $\left\langle \nabla F(\omega^{t}), \Delta_{m}^{t}\right\rangle$ as follows
	\begin{equation*}
	\begin{split}
		&\mathbb{E}\left\langle \nabla F(\omega^{t}), \Delta_{m}^{t}\right\rangle\\
		&=\underbrace{\mathbb{E}\left\langle \nabla F(\omega^{t}), \nabla f_{m}({\omega}^{t}; \xi_{m}^{t})\!-\!\nabla f_{m}({\omega}^{t-\tau_{m}^{t}}; \xi_{m}^{t})\right\rangle}_{\uppercase\expandafter{\romannumeral1}} \\
		&\quad +\underbrace{\mathbb{E}\left\langle \nabla F(\omega^{t}), \nabla f_{m}({\omega}^{t-\tau_{m}^{t}}; \xi_{m}^{t})\!-\!\nabla f_{m}({\omega}^{t-\tau_{m}^{t}}; \xi_{m}^{t-\tau_{m}^{t}})\right\rangle}_{\uppercase\expandafter{\romannumeral2}}.
	\end{split}
	\end{equation*}
	By using the inequality $\langle a, b \rangle \le \frac{\varepsilon}{2} \|a\|^{2}+\frac{1}{2 \varepsilon}\|b\|^{2}$ with $\varepsilon:=L\gamma$, and employing selection rule \eqref{condition}, we obtain 
	\begin{equation*}
	\uppercase\expandafter{\romannumeral1} \leq \frac{L\gamma}{2}\mathbb{E}[\|\nabla F(\omega^{t})\|^{2}]+\frac{1}{2L\gamma}\frac{1}{M^{2}}\sum_{d=1}^{D}\alpha_{d}\|\omega^{t+1-d}\!-\!\omega^{t-d}\|^{2}.
	\end{equation*}
	Noticing that
	\begin{equation*}
	\begin{split}
		\mathbb{E}\langle \nabla F(\omega^{t-\tau_{m}^{t}}), \nabla f_{m}({\omega}^{t-\tau_{m}^{t}}; \xi_{m}^{t})\!-\!\nabla f_{m}({\omega}^{t-\tau_{m}^{t}}; \xi_{m}^{t-\tau_{m}^{t}})\rangle\!=\!0,
	\end{split}
	\end{equation*}
	we then have
	\begin{equation*}
		\begin{split}
			\uppercase\expandafter{\romannumeral2}&\!\leq\! L\mathbb{E}\langle \omega^{t}\!-\!\omega^{t-\tau_{m}^{t}}\!,\!\nabla f_{m}({\omega}^{t-\tau_{m}^{t}}; \xi_{m}^{t})\!-\!\nabla f_{m}({\omega}^{t-\tau_{m}^{t}}; \xi_{m}^{t-\tau_{m}^{t}})\rangle\\
			& \!\leq\! \frac{\gamma DL}{2}\mathbb{E}\big[\|\nabla f_{m}({\omega}^{t-\tau_{m}^{t}}; \xi_{m}^{t})-\nabla f_{m}({\omega}^{t-\tau_{m}^{t}}; \xi_{m}^{t-\tau_{m}^{t}})\|^{2}\big] \\
			&\quad +\frac{L}{2\gamma D}\|\omega^{t}-\omega^{t-\tau_{m}^{t}}\|^{2}\\
			& \!\leq\! 2\gamma DL \sigma_{m}^{2}+\frac{L}{2\gamma}\sum_{d=1}^{D}\|\omega^{t+1-d}-\omega^{t-d}\|^{2},
		\end{split}
	\end{equation*}
	where we use the inequality $\langle a, b \rangle \le \frac{\varepsilon}{2} \|a\|^{2}+\frac{1}{2 \varepsilon}\|b\|^{2}$ with $\varepsilon:=D\gamma$, $\|\sum_{i=1}^{n} \theta_{i}\|^{2} \leq n \sum_{i=1}^{n}\|\theta_{i}\|^{2}$, and Assumption \ref{amp:2}. Then, we arrive at (with $\sigma^{2}:=\sum_{m=1}^{M}\sigma_{m}^{2}$)
	\begin{equation*}
	\begin{split}
		&\frac{\gamma}{M} \mathbb{E}\Big\langle \nabla F(\omega^{t}), \sum_{m \in \mathcal{M}_{c}^{t}} \Delta_{m}^{t}\Big\rangle \leq \frac{\gamma}{M} \sum_{m \in \mathcal{M}_{c}^{t}} \left(\uppercase\expandafter{\romannumeral1}+\uppercase\expandafter{\romannumeral2}\right)\\
		&\leq \frac{L\gamma^{2}}{2}\mathbb{E}\|\nabla F(\omega^{t})\|^{2}\!+\!\sum_{d=1}^{D}(\frac{\alpha_{d}}{2M^{2}L}\!+\!\frac{L}{2})\|\omega^{t+1-d}\!-\!\omega^{t-d}\|^{2}\\
		&\quad +\frac{2\gamma^{2} DL \sigma^{2}}{M}.
	\end{split}
	\end{equation*}
	By further analyzing $\|\Delta_{m}^{t}\|^2$ and $\|\nu^{t+1}-\nu^{t}\|^{2}$, we derive $(e)$
	\begin{equation}
	\label{delta_ana}
	\begin{split}
		&\|\sum_{m \in \mathcal{M}_{c}^{t}}\Delta_{m}^{t}\|^{2} \leq |\mathcal{M}_{c}^{t}|\sum_{m \in \mathcal{M}_{c}^{t}}\|\Delta_{m}^{t}\|^2\\
		&\leq |\mathcal{M}_{c}^{t}|\sum_{m \in \mathcal{M}_{c}^{t}}\Big\|\nabla f_{m}({\omega}^{t}; \xi_{m}^{t})-\nabla f_{m}({\omega}^{t-\tau_{m}^{t}}; \xi_{m}^{t})\\
		&\quad+\nabla f_{m}({\omega}^{t-\tau_{m}^{t}}; \xi_{m}^{t})-\nabla f_{m}({\omega}^{t-\tau_{m}^{t}}; \xi_{m}^{t-\tau_{m}^{t}})\Big\|^{2}\\
		&\leq 2M\sum_{m \in \mathcal{M}_{c}^{t}}\|\nabla f_{m}({\omega}^{t}; \xi_{m}^{t})-\nabla f_{m}({\omega}^{t-\tau_{m}^{t}}; \xi_{m}^{t})\|^{2}\\
		&\quad +2M\sum_{m \in \mathcal{M}_{c}^{t}}\|\nabla f_{m}({\omega}^{t-\tau_{m}^{t}}; \xi_{m}^{t})\!-\!\nabla f_{m}({\omega}^{t-\tau_{m}^{t}}; \xi_{m}^{t-\tau_{m}^{t}})\|^{2}\\
		& \leq 2\sum_{d=1}^{D}\alpha_{d}\left\|\omega^{t+1-d}-\omega^{t-d}\right\|^{2}+8M\sigma^{2},
	\end{split}
	\end{equation}
	where $\sigma^{2}:=\sum_{m=1}^{M}\sigma_{m}^{2}$. Applying inequalities \eqref{aux:appv}, \eqref{delta_ana}, and Assumption \ref{amp:2}, we can get
	\begin{equation}
	\label{vv}
	\begin{split}
		&\mathbb{E}[\|\nu^{t+1}-\nu^{t}]\|^{2}\\
		&\!=\!\mathbb{E}[\|-\frac{\gamma}{M} \sum_{m \in \mathcal{M}}\nabla f_{m}({\omega}^{t}; \xi_{m}^{t})+\frac{\gamma}{M} \sum_{m \in \mathcal{M}_{c}^{t}}\Delta_{m}^{t}\|^{2}]\\
		&\!\leq\! \frac{2\gamma^2}{M^{2}}\mathbb{E}[\|\sum_{m \in \mathcal{M}}\nabla f_{m}({\omega}^{t}; \xi_{m}^{t})\|^{2}]+\frac{2\gamma^2}{M^{2}}\mathbb{E}[\|\sum_{m \in \mathcal{M}_{c}^{t}}\Delta_{m}^{t}\|^{2}]\\
		&\!\leq\! 4\gamma^2\mathbb{E}\|\nabla F(\omega^{t})\|^{2}\!+\!\frac{4\gamma^2}{M^{2}}\sum_{d=1}^{D}\alpha_{d}\|\omega^{t+1-d}\!-\!\omega^{t-d}\|^{2}\!+\!\frac{20\gamma^{2}\sigma^{2}}{M}
	\end{split}
	\end{equation}
	Organize and summarize these items, and we can get the final result \eqref{F_descent}. 

	\subsection{Proof of Lemma \ref{lem:4}}
	\label{app:a-5}
	\noindent According to the definition \eqref{lya} and Lemma \ref{lem:3}, a straightforward calculation yields
	\begin{equation*}
	\begin{split}
		&\mathcal{L}^{t+1}-\mathcal{L}^{t}=\mathbb{E}[F(\nu^{t+1})]-\mathbb{E}[F(\nu^{t})]\\
		&\quad +\sum_{d=1}^{D}\beta_{d}\|\omega^{t+2-d}\!-\!\omega^{t+1-d}\|^{2}-\sum_{d=1}^{D}\beta_{d}\|\omega^{t+1-d}\!-\!\omega^{t-d}\|^{2}
	\end{split}
	\end{equation*}
	\begin{equation*}
	\begin{split}
		&\leq -(\gamma-\frac{5L+4}{2}\gamma^2)\mathbb{E}\|\nabla F({\omega}^{t})\|^{2}+\frac{L^{2}}{2M^{2}}\mathbb{E}\|\sum_{m \in \mathcal{M}}e_{m}^{t}\|^{2}\\
		&\quad +\sum_{d=1}^{D}\left[\frac{\alpha_{d}}{2M^{2}L}+\frac{L}{2}+\frac{2\gamma^2(L\!+\!1)\alpha_{d}}{M^{2}}\right]\|\omega^{t+1-d}\!-\!\omega^{t-d}\|^{2}\\
		&\quad +\left[10(L+1)+2DL\right]\frac{\gamma^{2}\sigma^{2}}{M} +\beta_{1}\|\omega^{t+1}-\omega^{t}\|^{2}\\
		&\quad+\sum_{d=1}^{D-1}(\beta_{d+1}-\beta_{d})\|\omega^{t+1-d}-\omega^{t-d}\|^{2}\\
		&\quad -\beta_{D}\|\omega^{t+1-D}-\omega^{t-D}\|^{2}.
	\end{split}
	\end{equation*}
	By utilizing the properties \eqref{aux:appv}, \eqref{vv}, and also invoking Lemma \ref{lem:2}, we can establish the following inequality
	\begin{equation*}
	\begin{split}
		\|\omega^{t+1}-\omega^{t}\|^{2}&\leq 2\|\nu^{t+1}-\nu^{t}\|^{2}+\frac{2}{M^{2}}\|\sum_{m \in \mathcal{M}}(e_{m}^{t+1}-e_{m}^{t})\|^{2}\\
		&\leq 2\|\nu^{t+1}-\nu^{t}\|^{2}+\frac{16(1-\delta)\gamma^{2} B^{2}}{\delta^{2}M},
	\end{split}
	\end{equation*}
	where $B^{2}:=\sum_{m=1}^{M}B_{m}^{2}$. We then derive that 
	\begin{equation*}
	\begin{split}
		\mathcal{L}^{t+1}-\mathcal{L}^{t}&\leq -\Big[\gamma-(\frac{5L+4}{2}+8\beta_{1})\gamma^2\Big]\mathbb{E}[\|\nabla F({\omega}^{t})\|^{2}] \\
		&\quad -\sum_{d=1}^{D}c_{d}\|\omega^{t+1-d}\!-\!\omega^{t-d}\|^{2}+a\frac{\gamma^{2}\sigma^{2}}{M}+b\frac{\gamma^{2}B^{2}}{M},
	\end{split}
	\end{equation*}
	where
	\begin{equation}
	\label{para}
	\left\{
	\begin{array}{l}
		\displaystyle
		c_{f}:=\gamma-(\frac{5L+4}{2}+8\beta_{1})\gamma^2 \vspace{1.0ex}\\
		\displaystyle
		c_{d}:=\beta_{d}\!-\!\beta_{d+1}\!-\!\left(\frac{\alpha_{d}}{2M^{2}L}\!+\!\frac{L}{2}\!+\!\frac{2\gamma^2\alpha_{d}(L\!+\!1\!+\!4\beta_{1})}{M^{2}}\right) \vspace{1.0ex}\\
		\qquad  d=1, \ldots, D-1 \vspace{1.1ex}\\
		\displaystyle
		c_{D}:=\beta_{D}-\left(\frac{\alpha_{D}}{2M^{2}L}+\frac{L}{2}+\frac{2\gamma^2\alpha_{D}(L+1+4\beta_{1})}{M^{2}}\right) \vspace{1.1ex}\\
		\displaystyle
		a:=10(L+1)+2DL+40\beta_{1}  \vspace{1.1ex}\\
		\displaystyle
		b:=\frac{(2L^{2}+16\beta_{1})(1-\delta)}{\delta^{2}}.
	\end{array}
	\right.
	\end{equation}

	\subsection{Proof of Theorem \ref{thm:1}}
	\label{app:a-6}
	\noindent Let $\gamma \leq \overline{\gamma}=\frac{1}{5L+4+16\beta_{1}}$, then choose $\{\beta_{d}\}_{d=1}^{D}$ such that
	\begin{equation*}
	\left\{
	\begin{array}{l}
		\displaystyle
		\beta_{d}\!-\!\beta_{d+1}\!-\!\left[\frac{\alpha_{d}}{2M^{2}L}\!+\!\frac{L}{2}\!+\!\frac{2\overline{\gamma}^2\alpha_{d}(L\!+\!1\!+\!4\beta_{1})}{M^{2}}\right]=0\\
		\qquad  d=1, \ldots, D-1 \\
		\displaystyle
		\beta_{D}-\left[\frac{\alpha_{D}}{2M^{2}L}+\frac{L}{2}+\frac{2\overline{\gamma}^2\alpha_{D}(L+1+4\beta_{1})}{M^{2}}\right]=0.\\
	\end{array}
	\right.
	\end{equation*}
	Solving the linear equations above we can further get
	\begin{equation*}
	\beta_{1}=\frac{[\frac{2\overline{\gamma}^{2}(L+1)}{M^{2}}+\frac{1}{2M^{2}L}]\sum_{d=1}^{D}\alpha_{d}+\frac{LD}{2}}{1-\frac{8\overline{\gamma}^{2}}{M^{2}}\sum_{d=1}^{D}\alpha_{d}}.
	\end{equation*}
	We then have $c_{d}\geq 0, d=1, \ldots, D$ and $c_{f}\geq\gamma/2$, implying
	\begin{equation*}
	\mathcal{L}^{t+1}-\mathcal{L}^{t}\leq -\frac{\gamma}{2}\mathbb{E}[\|\nabla F({\omega}^{t})\|^{2}]+\frac{a\sigma^{2}+bB^{2}}{M}\gamma^{2}.
	\end{equation*}
	By taking the summation $\sum\limits_{t=0}^{T-1}\mathcal{L}^{t+1}-\mathcal{L}^{t}$, we arrive at
	\begin{equation*}
		\begin{split}
			\sum_{t=0}^{T-1}\frac{\gamma}{2}\mathbb{E}[\|\nabla F({\omega}^{t})\|^{2}] \leq F(\omega^{0})-F^{*}+\frac{a\sigma^{2}+bB^{2}}{M}T\gamma^{2}.
		\end{split}
	\end{equation*}
	If we choose learning rate $\gamma\!=\!\min \{1/(5L\!+\!4\!+\!16\beta_{1}), c_{\gamma}/\sqrt{T}\}$, where $c_{\gamma}>0$ is a constant. We then have
	\begin{equation}
	\nonumber
	\begin{split}
		&\frac{1}{T}\sum_{t=0}^{T\!-\!1}\mathbb{E}[\|\nabla F({\omega}^{t})\|^{2}] \!\leq\! \frac{2M(F(\omega^{0})\!-\!F^{*})\!+\!2T(a\sigma^{2}\!+\!bB^{2})\gamma^{2}}{MT\gamma}\\
		&\leq \frac{2(5L+4+16\beta_{1})(F(\omega^{0})-F^{*})}{T}+\frac{2(F(\omega^{0})-F^{*})}{c_{\gamma}\sqrt{T}} \\
		&\quad +\frac{2c_{\gamma}(a\sigma^{2}+bB^{2})}{M\sqrt{T}},
	\end{split}
	\end{equation}
	which means
	$$
	\frac{1}{T}\sum_{t=0}^{T-1}\mathbb{E}\left[\|\nabla F({\omega}^{t})\|^{2}\right]=\mathcal{O}(1/\sqrt{T}).
	$$

	\section{Experiment results}
	\label{sec:exp}
	
	This section conducts extensive experiments to demonstrate the convergence properties and communication efficiency of our SASG algorithm. The experiments encompass simulated and real distributed training scenarios, examining the improvement of SASG in both communication volume and time. Additionally, we discuss the additional computational and memory overhead of SASG, the effect of sparsity rate, and the algorithm's performance in heterogeneous data scenarios.

	\begin{figure*}[t]
		\centering
		\subfloat[\footnotesize FC+MNIST]{
			\label{test:a}
			\includegraphics[width=0.3\linewidth]{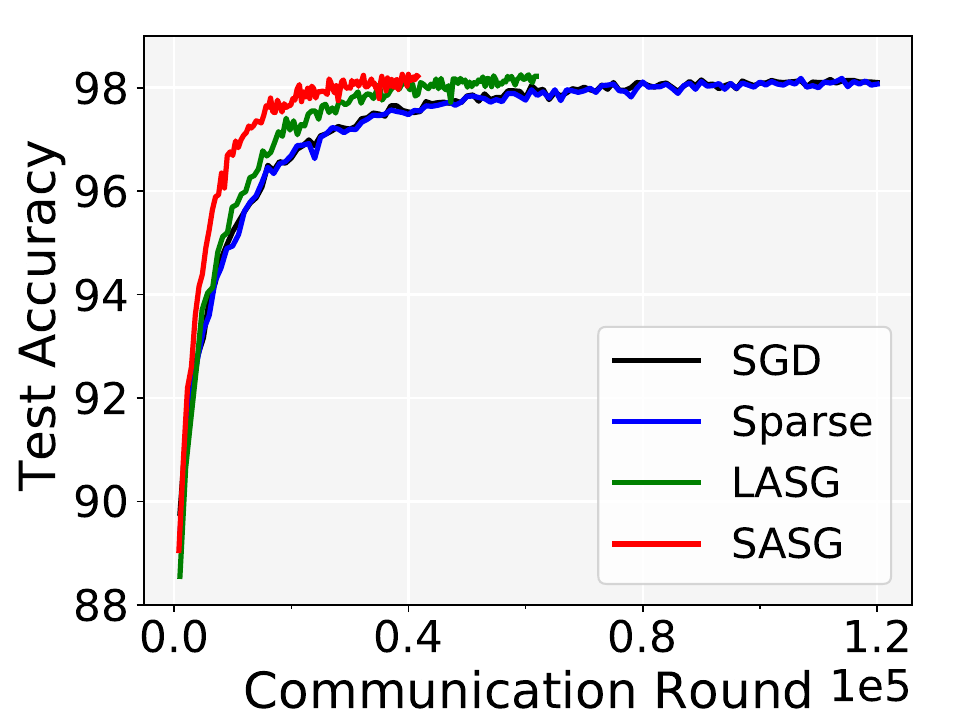}}
		\subfloat[\footnotesize ResNet18+CIFAR-10]{
			\label{test:b}
			\includegraphics[width=0.3\linewidth]{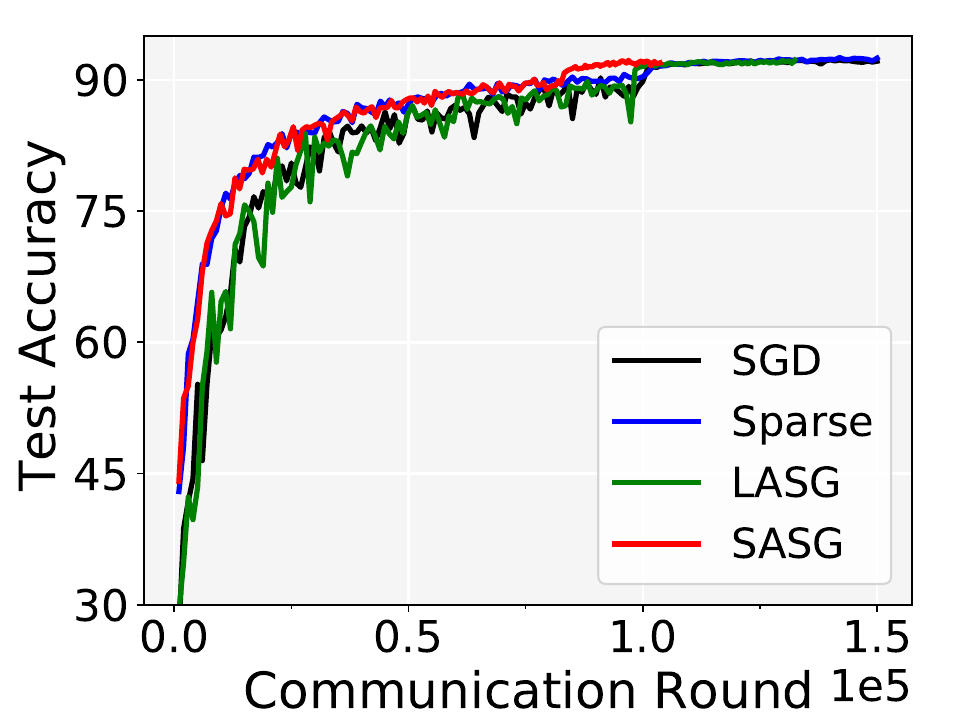}}
		\subfloat[\footnotesize ResNet18+CIFAR-100]{
			\label{test:c}
			\includegraphics[width=0.3\linewidth]{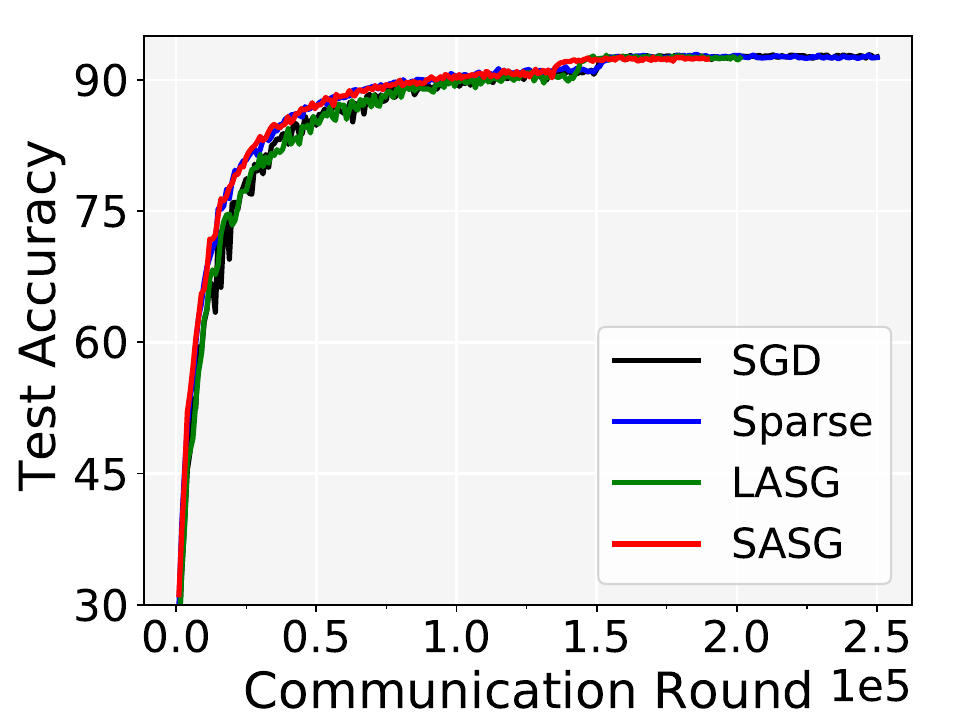}}
		\\
		\subfloat[\footnotesize FC+MNIST]{
			\label{loss:a}
			\includegraphics[width=0.3\linewidth]{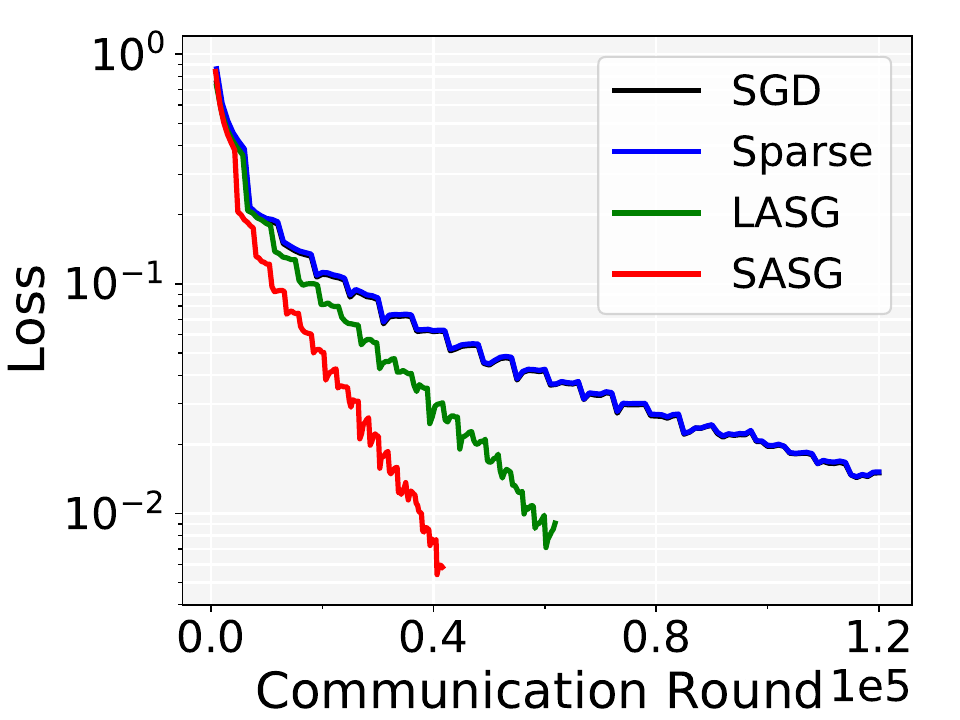}}
		\subfloat[\footnotesize ResNet18+CIFAR-10]{
			\label{loss:b}
			\includegraphics[width=0.3\linewidth]{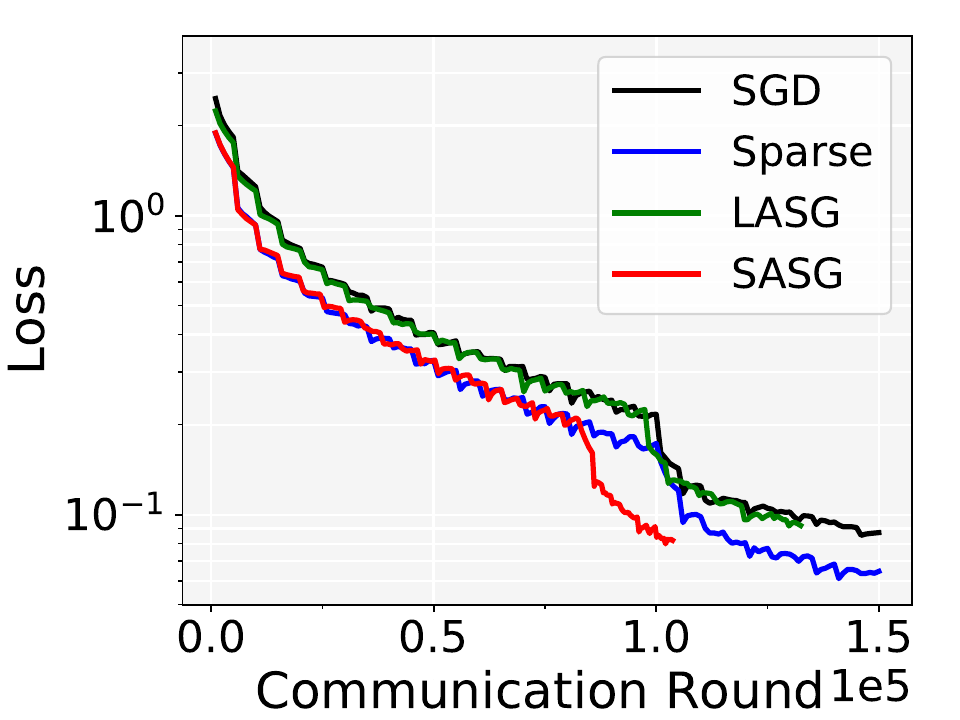}}
		\subfloat[\footnotesize ResNet18+CIFAR-100]{
			\label{loss:c}
			\includegraphics[width=0.3\linewidth]{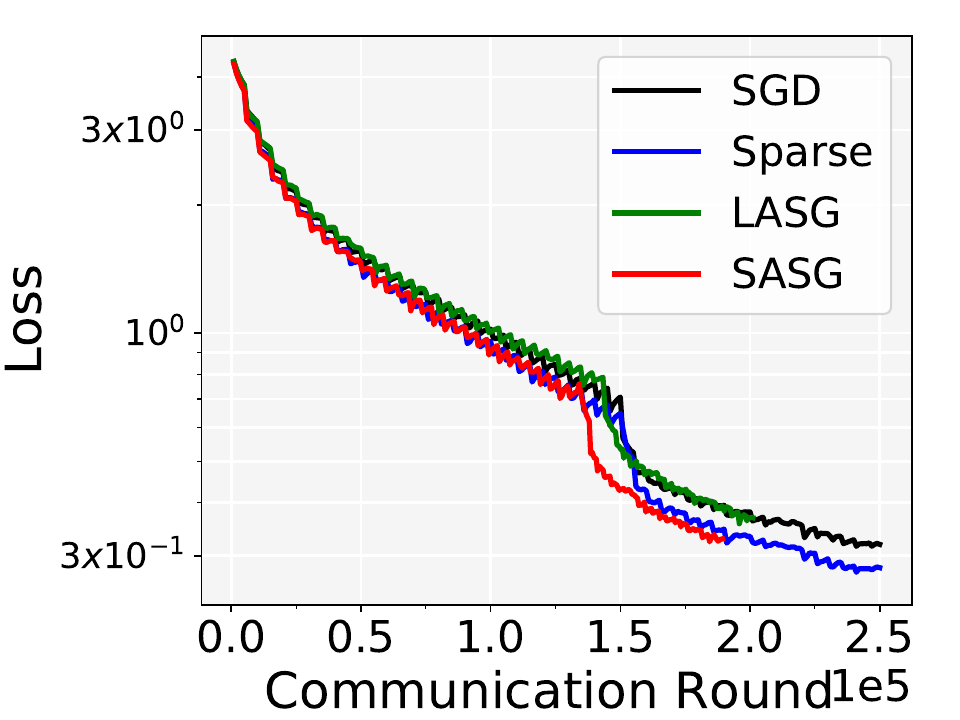}}
		\caption{Experimental results of test accuracy percentage and training loss versus communication round in three settings. All four methods are trained with the same number of epochs. Our algorithm significantly reduces the number of communication rounds required to achieve the same performance and complete the training.}
		\label{fig:test+loss}
	\end{figure*}

	\subsection{Setup}
	We evaluate the performance of the SASG algorithm against LASG \cite{chen2020lasg}, the sparsification method with error feedback \cite{aji2017sparse}, and distributed SGD \cite{zinkevich2010parallelized}. Aji et al. \cite{aji2017sparse} demonstrated that up to 99\% of the gradients are not needed to update the model at each iteration. Hence, we utilize the top-1\% (i.e., $k=0.01d$) sparsification operator in SASG and the sparsification method. All our experimental results are based on the PyTorch framework \cite{PaszkeGMLBCKLGA19}.

	We begin by evaluating the communication volume of the four algorithms. We simulated ten workers, and each worker used ten samples per training iteration. Subsequently, we recorded the communication time for each algorithm in the real-world distributed settings. We employed ten Nvidia RTX-3090 GPUs as distributed workers for training, with the first GPU also playing the role of parameter server. During information uploads from the workers to the server, we utilized point-to-point communication with the GLOO backend by invoking the $\mathsf{send()}$ and $\mathsf{recv()}$ functions in PyTorch. We configured the upload bandwidth of each worker to be 1 Gbps. We carried out evaluations for the following three settings and repeated every experiment five times.

	\smallskip
	\noindent{\textbf{MNIST:}} The MNIST \cite{lecun1998gradient} dataset contains 70,000 handwritten digits in 10 classes, with 60,000 examples in the training set and 10,000 examples in the test set. We consider a two-layer fully connected (FC) neural network model with 512 neurons in the second layer for 10-category classification on MNIST. For all algorithms, we trained $20$ epochs with the learning rate $\gamma=0.005$. For the adaptive aggregated algorithms SASG and LASG, we set $D = 10, \alpha_{d} = 1/2\gamma$ for $d = 1, 2, \ldots, 10$.
	
	\smallskip
	\noindent{\textbf{CIFAR-10:}} The CIFAR-10 \cite{krizhevsky2009learning} dataset comprises 60,000 colored images in 10 classes, with 6,000 images per category. We train ResNet18 \cite{he2016deep} with all the algorithms mentioned above on the CIFAR-10 dataset. Standard data augmentation techniques such as random cropping, flipping, and normalization are performed. We trained $30$ epochs for the four algorithms, and the basic learning rate is set to $\gamma = 0.01$, with a learning rate decay of $0.1$ at epoch $20$. For SASG and LASG, we set $D = 10, \alpha_{d} = 1/\gamma$ for $d = 1, 2, \ldots, 10$.
	
	\smallskip
	\noindent{\textbf{CIFAR-100:}} We also train ResNet18 on the CIFAR-100 \cite{krizhevsky2009learning} dataset, which consists of 60,000 images in 100 classes. A data augmentation technique similar to CIFAR-10 was performed. The basic learning rate is set to $\gamma = 0.01$, with a learning rate decay of $0.1$ at epoch $30$, where the total training epoch is $50$. In the case of SASG and LASG, we set $D = 10, \alpha_{d} = 1/\gamma$ for $d = 1, 2, \ldots, 10$.

	It is important to note that we did not employ training techniques such as warm-up or weight decay in our experiments, and thus state-of-the-art performance was not achieved in certain models. Our goal is to demonstrate, through comparative experiments, that the SASG algorithm can effectively reduce both the number of communication rounds and bits without sacrificing model performance.

	\begin{figure*}[t]
		\centering
		\subfloat[\footnotesize Accuracy v.s. Time]{
			\label{res_acc:a}
			\includegraphics[width=0.28\linewidth]{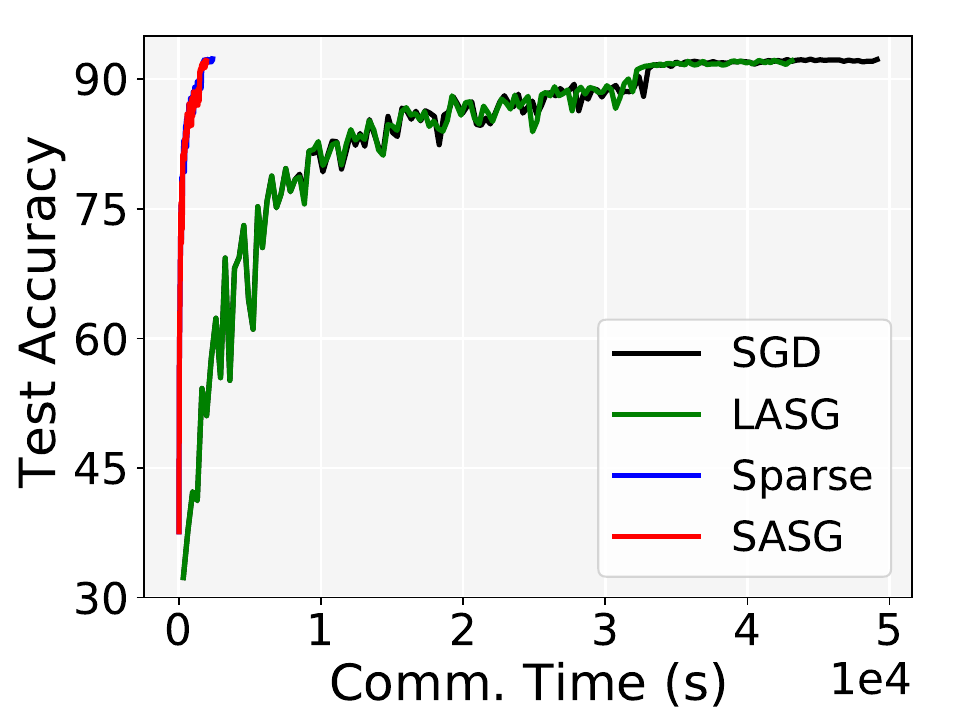}}
		\quad
		\subfloat[\footnotesize Accuracy v.s. Time]{
			\label{res_acc:b}
			\includegraphics[width=0.28\linewidth]{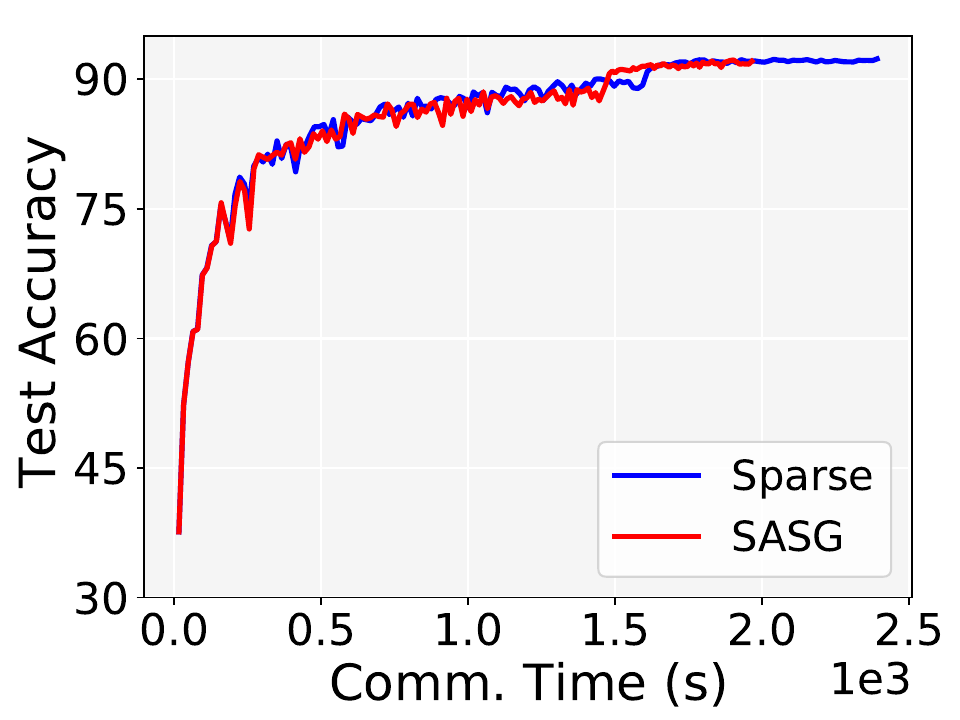}}
		\quad
		\subfloat[\footnotesize Time v.s. Iteration]{
			\label{res_acc:c}
			\includegraphics[width=0.28\linewidth]{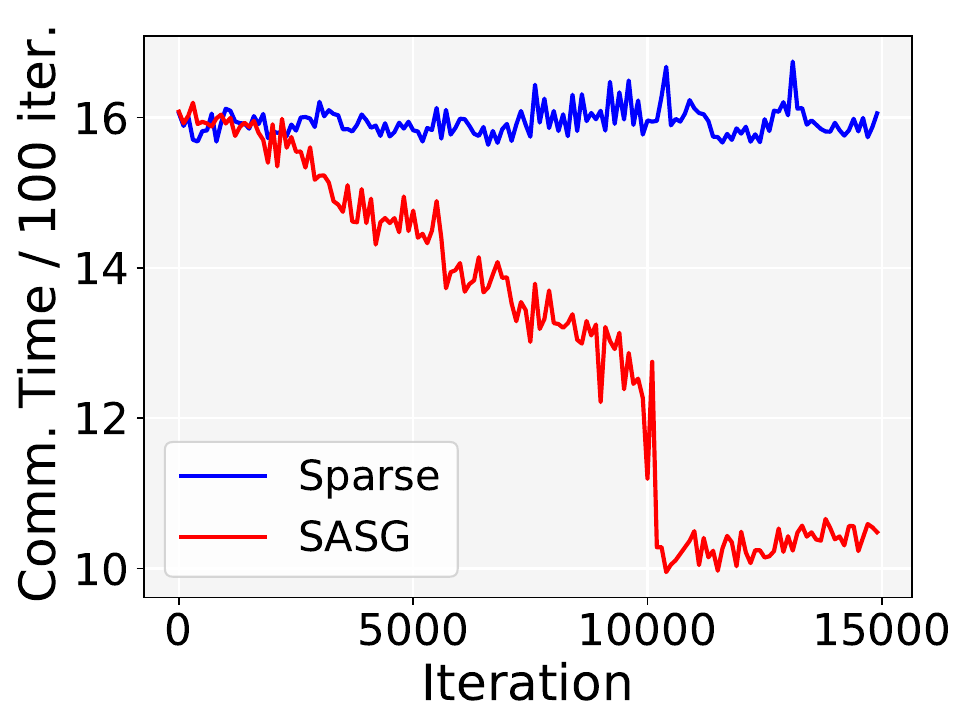}}
		\caption{Experimental results of classifying CIFAR-10 with ResNet18. Figures (a) and (b) show the test accuracy versus communication time. Figure (c) records the communication time every $100$ iterations, where the learning rate decay was used at the $10,000$-th iteration.}
		\label{fig:res}
	\end{figure*}

	\subsection{Communication volume}
	This part presents the results of our simulation experiments on communication volume, i.e., the number of communication rounds and bits. Figure \ref{fig:test+loss} depicts the variation of test accuracy and training loss against the number of communication rounds, whereas Figure \ref{test:c} shows the top-5 test accuracy.
	
	The experimental results demonstrate that our SASG algorithm outperforms the previous methods. Notably, as evident in Figure \ref{test:a}, SASG achieves higher test accuracy with the same number of communication rounds, indicating that our algorithm selects more valuable rounds for communication. Furthermore, SASG can significantly reduce the number of communication rounds while preserving the same model performance after completing the training process. Figure \ref{loss:a} shows that the SASG algorithm achieves faster and better convergence results with fewer communication rounds. In the experiments on the CIFAR-10 and CIFAR-100 datasets (Figure \ref{loss:b}-\ref{loss:c}), SASG is also able to significantly reduce the number of communication rounds required to complete the training while guaranteeing convergence.

	\begin{table}[t]
		\begin{center}
			{\caption{Number of communication rounds and bits required for the four algorithms to reach the same test accuracy baseline (average of five experiments).}
			\label{tab:bits}}
			\begin{tabular}{ccrc}
				\toprule[1pt]
				\rule{0pt}{10pt}
				Model \& Dataset & Method & \# Rounds& $~~~$\# Bits$~~~$ \\
				\midrule[0.5pt]
				\rule{0pt}{10pt}
				\multirow{4}{*}{\makecell[c]{FC \\ MNIST}} & SGD & 60400 $~$ & 7.87 E{+11}\\
				& Sparse & 67400 $~$ & 8.78 E{+09}\\
				& LASG & 33853 $~$ & 4.41 E{+11}\\
				& SASG & \textbf{22823} $~$ & \textbf{2.97 E{+09}}\\
				\midrule[0.5pt]
				\rule{0pt}{10pt}
				\multirow{4}{*}{\makecell[c]{ResNet18 \\ CIFAR-10}} & SGD & 111200 $~$ & 3.98 E{+13}\\
				& Sparse & 109800 $~$ & 3.93 E{+11}\\
				& LASG & 107389 $~$ & 3.84 E{+13}\\
				& SASG & \textbf{90846} $~$ & \textbf{3.25 E{+11}}\\
				\midrule[0.5pt]
				\rule{0pt}{10pt}
				\multirow{4}{*}{\makecell[c]{ResNet18 \\ CIFAR-100}} & SGD & 152000 $~$ & 5.46 E{+13}\\
				& Sparse & 152800 $~$ & 5.49 E{+11}\\
				& LASG & 145971 $~$ & 5.24 E{+13}\\
				& SASG & \textbf{137937} $~$ & \textbf{4.95 E{+11}}\\
				\bottomrule[1pt]
			\end{tabular}
		\end{center}
	\end{table}
	
	According to Figure \ref{fig:test+loss}, we choose several baselines (around the best accuracy) to present the specific number of communication rounds and bits required to achieve the same performance for the four algorithms, with the accuracy baseline set as 98\% for MNIST, 92\% for CIFAR-10 and CIFAR-100 (top-5 accuracy), respectively. As shown in Table \ref{tab:bits}, thanks to the adaptive aggregation technique, both SASG and LASG algorithms reduce the number of communication rounds, while our SASG algorithm is more effective. The number of communication bits required by the different algorithms to reach the same baseline can be obtained by calculating the number of parameters for different models. The last column of Table \ref{tab:bits} shows that the SASG algorithm, blending adaptive aggregation techniques with sparse communication, outperforms the LASG and sparsification methods, significantly reducing the number of communication bits required for the model to achieve the desired performance.
	
	\begin{table}[t]
		\begin{center}
			{\caption{Average communication time of the four algorithms and extra overhead required for the two adaptive aggregated methods when classifying CIFAR-10 with ResNet18 (record every 100 training iterations).}
			\label{tab:time}}
			\begin{tabular}{cccc}
				\toprule[1pt]
				\rule{0pt}{10pt}
				Method & Communication & Computation & Memory \\
				\midrule[0.5pt]
				SGD & 327.45s & --- & --- \\
				Sparse & $~$15.94s & --- & --- \\
				LASG & 286.72s & 1.25s & 426.25MB \\
				SASG & $~$\textbf{13.09}s & \textbf{1.25}s & $~~$\textbf{4.26}MB \\
				\bottomrule[1pt]
			\end{tabular}
		\end{center}
	\end{table}

	\subsection{Communication time}
	This part presents the communication time required for each algorithm in real-world distributed training. As mentioned in Section \ref{sec:algo}, the top-$k$ sparsification method can significantly reduce the communication bits. Thus, the sparsification method and the SASG algorithm require much less communication time than the LASG and SGD algorithms when completing the same training task, as shown in Figure \ref{res_acc:a}. The two methods utilizing the sparsification technique are compared in Figure \ref{res_acc:b}, which illustrates that SASG accomplishes the required accuracy faster and completes training in less communication time. Figure \ref{res_acc:c} recorded the communication time required by the sparsification method and the SASG algorithm for every $100$ training iterations. As training proceeds, SASG skips more communication rounds, leading to a gradual decrease in required communication time, while the sparsification method remains the same. Besides, Figure \ref{res_acc:b} implies that skipping redundant information does not compromise the final performance of the model, which also indicates that our adaptive selection criterion is effective. The average communication time is shown in Table \ref{tab:time}.

	\subsection{More experimental discussion}

	While the aforementioned experimental results show that SASG can significantly improve the communication efficiency while guaranteeing algorithm performance, we also need to discuss the additional overhead associated with implementing this algorithm, i.e., the extra computation time for calculating the selection rule and the memory overhead required for storing the stale gradients. As outlined in Algorithm \ref{alg:sasg}, in a single SASG iteration, each worker needs to store one duplicate of the previous model parameters to compute the auxiliary gradient, and the server needs to store the old gradient (sparse one) from each worker. Table \ref{tab:time} presents the additional overhead required for the two adaptive aggregated methods. The computational overhead is approximately the same for both methods, mainly for computing an auxiliary gradient. However, this overhead is negligible when compared to the communication time. On the other hand, the SASG algorithm only needs to store the sparsified gradients on the server side compared to LASG, significantly decreasing the additional memory overhead.
	
\begin{figure}[t]
		\centering
		\subfloat{
			\label{sparse_acc}
			\includegraphics[width=0.495\linewidth]{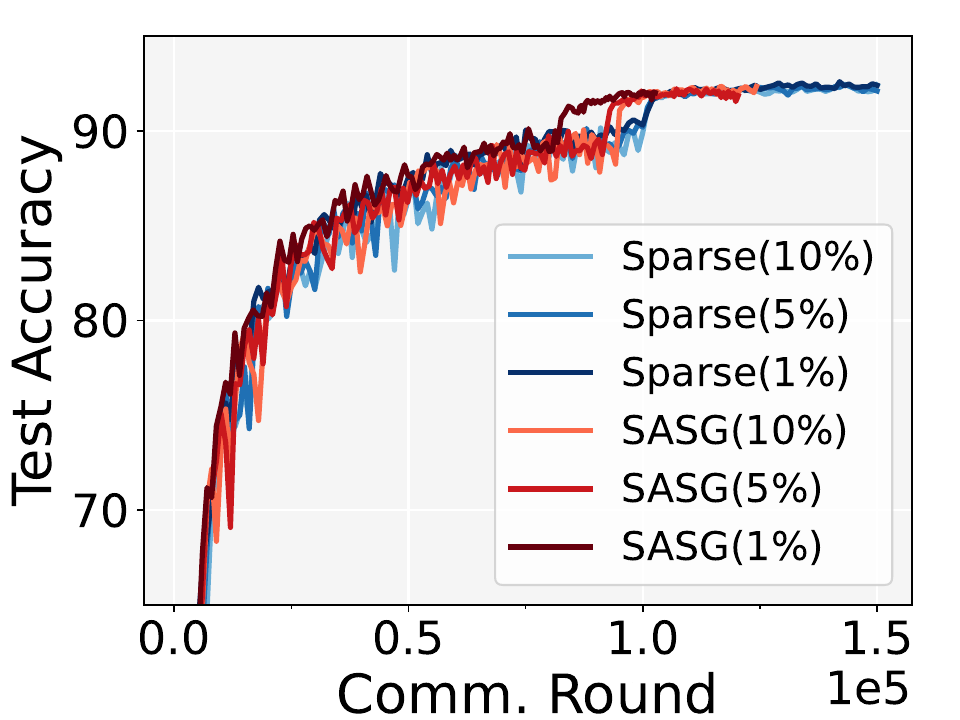}}
		\subfloat{
			\label{sparse_loss}
			\includegraphics[width=0.495\linewidth]{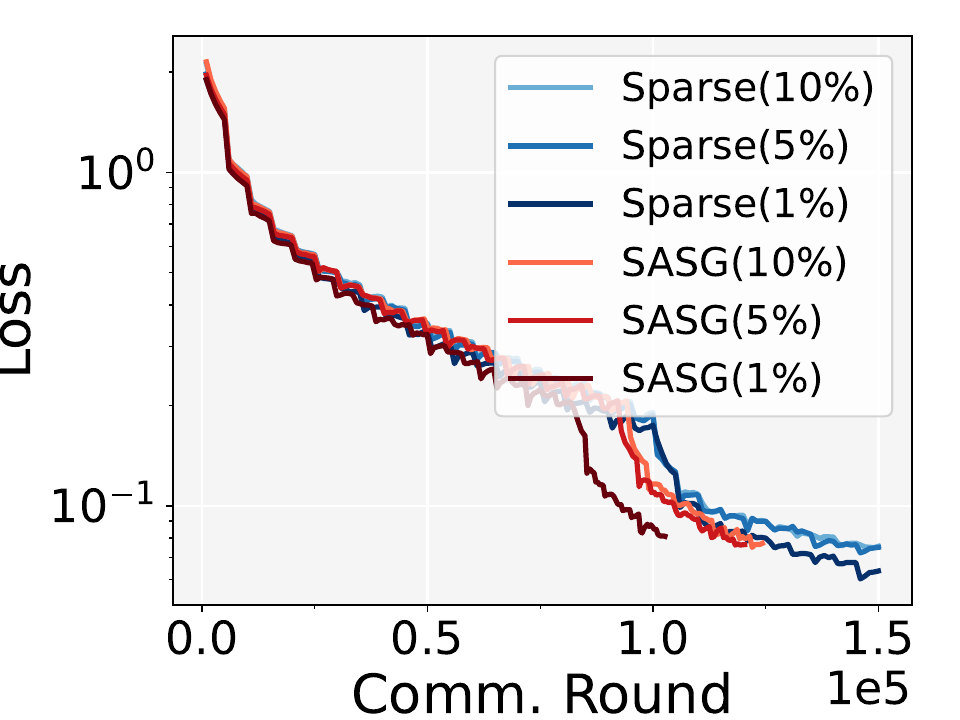}}
		\caption{Experimental results of training ResNet18 to classify CIFAR-10 using different sparsity rates.}
		\label{fig:sparse}
	\end{figure}

	Additionally, we explore the effect of different sparsification levels on SASG, as shown in Figure \ref{fig:sparse}. The results indicate that varying only the sparsification coefficient $k$ has no noticeable effect on the model accuracy and loss, and SASG still outperforms the sparsification method with error feedback. Besides, the experimental results reveal that SASG skips more communication rounds as the sparsity rate increases. Therefore, varying the sparse coefficient $k$ requires a corresponding adjustment of the parameter $\{\alpha_{d}\}_{d=1}^{D}$ within the adaptive aggregation selection criterion to prevent extreme sparsification from compromising the algorithm's performance. Moreover, the results in Figures \ref{fig:test+loss} and \ref{fig:sparse} show that the sparsification method with error feedback works better than the benchmark SGD algorithm, and its performance can be further improved by appropriately increasing the sparsification factor. The potential reason for this phenomenon can be attributed to the fact that the sparsification operation reduces the gradient norm (or the gradient upper bound), thereby improving the convergence \cite{aji2017sparse,stich2018sparsified} and generalization \cite{hardt2016train,sun2021stability,deng2023stability} of the algorithm.

\begin{figure}[t]
		\centering
		\subfloat{
			\label{noiid_acc}
			\includegraphics[width=0.495\linewidth]{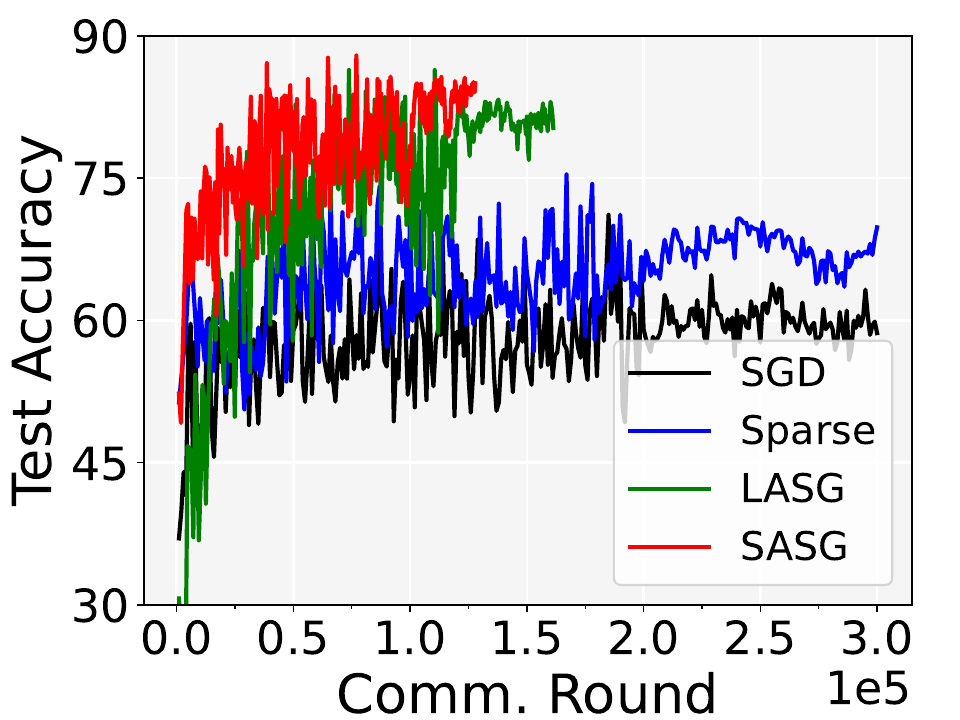}}
		\subfloat{
			\label{noiid_loss}
			\includegraphics[width=0.495\linewidth]{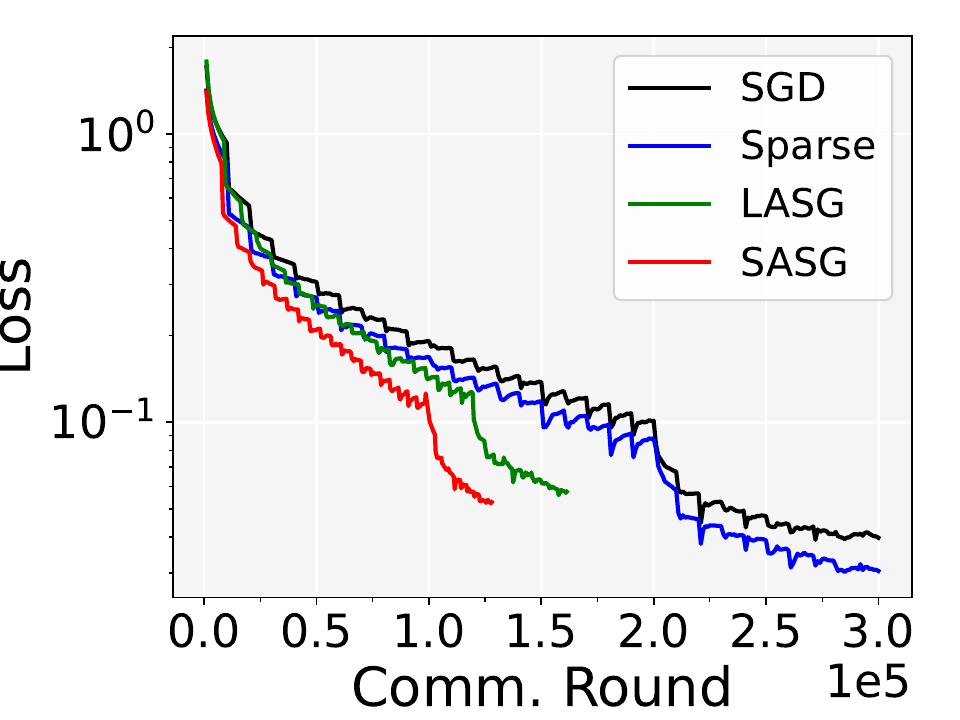}}
		\caption{Experimental results of training the ResNet18 model in heterogeneous data scenarios.}
		\label{fig:noniid}
	\end{figure}

	For the heterogeneous data experiments, we simulated ten distributed workers and mixed CIFAR-10 and MNIST datasets to construct the non-i.i.d. environment. Specifically, the even-numbered workers distribute only the CIFAR-10 dataset, while the odd-numbered workers contain only MNIST training samples. The experimental results of training the ResNet18 model in this non-i.i.d. setting using the four algorithms are depicted in Figure \ref{fig:noniid}. This result shows that the SASG algorithm can significantly improve the communication efficiency of the training system and ensure good model accuracy even in heterogeneous data scenarios. In addition, this experiment also demonstrates that judiciously skipping some communication rounds can compensate for the adverse impact on model accuracy caused by training with non-i.i.d. data (e.g., federated learning), which is consistent with the findings of CMFL \cite{DBLP:conf/icdcs/WangWL19}.
	
	\section{Conclusion}
	\label{sec:conclusion}
	This paper proposes a communication-efficient SASG algorithm for distributed learning. SASG adaptively skips several communication rounds with an adaptive selection rule, and further reduces the number of communication bits by sparsifying the transmitted information. For the biased nature of the top-$k$ sparsification operator, we utilize an error feedback framework and provide a sublinear convergence result for SASG with the help of a new Lyapunov analysis. Extensive experimental results show that SASG can reduce both the number of communication rounds and bits without sacrificing convergence performance, which corroborates our theoretical findings.

\bibliographystyle{IEEEtran}
\bibliography{tbd.bib}

\end{document}